%% file: mainPRD.tex
\documentclass[aps,prd,twocolumn,superscriptaddress,amsmath,preprintnumbers,amssymb,nofootinbib,longbibliography]{revtex4-2}

\usepackage{graphicx}
\usepackage[dvipsnames]{xcolor}
\usepackage{bbm}
\usepackage[colorlinks]{hyperref}
\usepackage[normalem]{ulem}

\usepackage{tikz}
\usetikzlibrary{tikzmark}
\usetikzlibrary{positioning}

\usepackage{physics}
\usepackage{xspace}
\usepackage{tabstackengine}
\usepackage{xparse}

\usepackage{siunitx}
\stackMath

\usepackage{aas_macros}

\hypersetup{linkcolor=BrickRed,citecolor=Green,
filecolor=Mulberry,
urlcolor=NavyBlue,
menucolor=BrickRed,
runcolor=Mulberry
}

\allowdisplaybreaks

\newcommand{\MESA}{\texttt{MESA}\xspace}
\newcommand{\Msun}{\ensuremath{{\rm M}_{\odot}}\xspace}
\newcommand{\MZAMS}{\ensuremath{{\rm M}_{\rm ZAMS}}\xspace}

\definecolor{lime}{HTML}{A6CE39}
\DeclareRobustCommand{\orcidicon}{\hspace{-1mm}
	\begin{tikzpicture}
	\draw[lime, fill=lime] (0,0) 
	circle [radius=0.16] 
	node[white] {{\fontfamily{qag}\selectfont \tiny \,ID}};
	\draw[white, fill=white] (-0.0525,0.095) 
	circle [radius=0.007];
	\end{tikzpicture}
	\hspace{-3mm}
}

\foreach \x in {A, ..., Z}{\expandafter\xdef\csname orcid\x\endcsname{\noexpand\href{https://orcid.org/\csname orcidauthor\x\endcsname}
			{\noexpand\orcidicon}}
}

\begin{document}

\title{The Effect of Mass Loss and Convective Overshooting on the Pre-Collapse Structure, Composition, and Neutrino Emission of Red Supergiants}

\input{authors}

\date{\today}

\begin{abstract}
Prior to core collapse, the neutrino emission from red supergiants (RSGs) is so large that a nearby ($\lesssim1\:$kpc) RSG will become visible in current and near-future neutrino detectors. The rate of emission and the spectra of the pre-supernova (pre-SN) neutrinos from RSGs are sensitive to the temperature, density, and detailed isotopic composition of the core. During the last year of the star's life, these properties change considerably as the nuclear burning accelerates and deleptonization begins. Uncertainties in stellar evolution modeling---including the treatment of mass loss and convective overshooting---alter the thermal conditions and composition of the RSG core as it approaches collapse, and thus one expects a consequent effect upon the pre-SN emission. 
In this paper we present the first study of how varying the treatment of mass loss and convective overshooting together affects the pre-collapse core properties and neutrino emission of RSGs. We use the stellar evolution instrument \MESA and construct a grid of 32 RSG models with zero-age main sequence masses of $\{ 12, 15, 18, 20\}$ \Msun, apply the so-called ``Dutch'' mass-loss scheme with wind efficiencies of $\{0.2, 0.4, 0.8, 1.0\}$, and consider two convective overshooting schemes. Our models use a large 206-isotope nuclear network in order to accurately compute the structure and composition of the star. We find that varying the treatment of mass loss and overshooting results in significant differences in core properties and the strength and timing of shell burning episodes, and that the onset of silicon shell burning leads to a localized peak in the pre-SN neutrino event rate observed by a JUNO-like detector $\sim 10$ hours to $\sim 1$ hour prior to collapse.

\end{abstract}

\maketitle


\input{bodyPRD}

\input{acknowledgments}

\bibliography{references}


\end{document}

%% file: authors.tex
\newcommand{\orcidauthorA}{OrcID: 0000-0002-2901-9173}
\author{McKenzie A. Myers\orcidA{}}
\affiliation{Department of Physics, NC State University, Raleigh, NC 27695, USA}

\newcommand{\orcidauthorB}{OrcID: 0009-0002-5773-3531}
\author{Claire B. Campbell\orcidB{}}
\affiliation{Department of Physics, Illinois State University, Normal, IL 61790, USA}

\newcommand{\orcidauthorC}{OrcID: 0000-0002-2154-4782}
\author{Kelly M. Patton\orcidC{}}
\affiliation{Department of Physics, Trinity College, Hartford, CT 06106, USA}

\newcommand{\orcidauthorD}{0000-0001-5537-4710}
\author{Segev BenZvi\orcidD{}}
\affiliation{Department of Physics and Astronomy, University of Rochester, 500 Wilson Blvd., Rochester, NY 14627, USA}

\newcommand{\orcidauthorE}{0000-0003-1801-8121}
\author{Marta Colomer Molla\orcidE{}}
\affiliation{Université Libre de Bruxelles, 1050 Bruxelles, Belgium}

\newcommand{\orcidauthorF}{0000-0002-1018-9383}
\author{Alec Habig\orcidF{}}
\affiliation{Department of Physics and Astronomy, University of Minnesota Duluth, 1023 University Dr., Duluth, MN 55812}

\newcommand{\orcidauthorG}{0000-0002-3502-3830}
\author{James P. Kneller\orcidG{}}
\email{jpknelle@ncsu.edu}
\affiliation{Department of Physics, NC State University, Raleigh, NC 27695, USA}

\newcommand{\orcidauthorH}{0000-0002-0763-3885}
\author{Dan Milisavljevic\orcidH{}}
\affiliation{Department of Physics and Astronomy, Purdue University, 525 Northwestern Ave., West Lafayette, IN 47907, USA}

\newcommand{\orcidauthorI}{0000-0003-1731-5853}
\author{Jeffrey Tseng\orcidI{}}
\affiliation{Department of Physics, University of Oxford, Oxford OX1 3RH, United Kingdom}

\collaboration{The SNEWS Collaboration}

%% file: bodyPRD.tex
\section{Introduction}

From pre-explosion images of Type IIP CCSNe, it is now clearly established that red supergiants (RSGs) are the final stage of the life of most massive stars before they explode \cite{2017ars..book.....L,2017RSPTA.37560270D,2025Galax..13...33V}. Should such an explosion occur within the Milky Way, our first indication of the event will be a pulse of neutrinos and gravitational waves passing across the Earth. The Supernova Early Warning System (SNEWS) has been established to look for the neutrino burst that would occur in terrestrial neutrino detectors and issue an alert that the supernova has begun \cite{2020NatRP...2..458H,2021NJPh...23c1201A,2024JInst..19P0017K}. Pre-explosion observations of whichever RSG in the Milky Way becomes the next supernova would aid greatly in connecting stellar evolution theory to the consequences, e.g., the amount of metals ejected by supernovae and the remnant mass. For this reason, a list of 640 RSGs in the Milky Way, $\sim 6\%$ of the estimated $\sim 11,000$ total \cite{2025A&A...694A.152Z}, has recently been compiled by Healy \emph{et al.} \cite{2024MNRAS.529.3630H} for the SNEWS collaboration, and the American Association of Variable Star Observers (AAVSO) is monitoring many stars on this list. 

Like all stars, RSGs emit neutrinos due to a combination of ``beta" and ``thermal" processes \cite{1967ApJ...150..979B,1996ApJS..102..411I,Farag_2020}. Beta processes, such as electron/positron capture or $\beta$-decay, are reactions which involve conversion of a proton into a neutron (or vice versa), whereas thermal processes---such as pair annihilation---do not. Only beta processes alter the electron fraction of the stellar material. After the star begins carbon burning---about 1,000 years before collapse---the neutrino luminosity of the RSG surpasses its electromagnetic luminosity by a factor that eventually exceeds $\sim 10^{10}$ before the star explodes \cite{RevModPhys.74.1015,Farag_2020}. The pre-supernova (pre-SN) neutrino emission from RSGs is so large that a nearby RSG will become visible in neutrino detectors here on Earth in the days to hours before it explodes \cite{ODRZYWOLEK2004303,yoshida_presupernova_2016,2016ApJ...818...91A,2019PhLB..796..126G,2020ARNPS..70..121K,2020ApJ...899..153M,2024ApJ...973..140A}. These emissions are a sensitive function of an RSG's core composition, density, temperature, and nuclear physics. It is apparent from previous studies that these properties of the RSG core are affected by significant uncertainties in stellar evolution theory. Accordingly, one should also expect the pre-SN neutrino emission to be affected.

One of the most significant uncertainties in stellar evolution is mass loss  \cite{2015A&A...575A..60M,2017A&A...603A.118R, 2025Galax..13...72V, 2021ARA&A..59..337D}, and mass-loss rates are a matter of some contention. Mass-loss rates are demonstrated to be proportional to luminosity, inversely proportional to effective temperature \cite{2024A&A...686A..88A}, and proportional to the RSG’s radius \cite{2024OJAp....7E..47F} but not strongly dependent upon metallicity \cite{2025A&A...702A.178A}. While the historically popular ``Dutch'' prescription \cite{1988A&AS...72..259D} assumes that mass-loss rates depend most heavily on luminosity and effective temperature, some more recent studies propose that it is the wind profile density distribution that most strongly affects mass-loss rates \cite{2024A&A...686A..88A}. Others have found evidence supporting the idea that mass-loss rates are strongly related to the convective velocity of RSG chromospheres \cite{2024OJAp....7E..47F}. Whatever the dependence, recent studies suggest that mass-loss rates used in stellar modeling should be kept low due to the rarity of observed Type Ic SNe and Wolf-Rayet stars that have been stripped almost entirely of their H envelopes; see, e.g., \cite{2020MNRAS.492.5994B}. 
Interestingly, while one might expect the choice of mass-loss treatment in the evolution modeling to affect the pre-SN surface properties of the star, Renzo \emph{et al.} \cite{2017A&A...603A.118R} found that varying the mass-loss rates did not appreciably affect the final position on the HR diagram except for stars with a zero-age main sequence (ZAMS) mass $\geq35\;\Msun$.

Another factor that will affect the structure and composition of the RSG, and thus the pre-SN emission, is overshooting at the convective-radiative boundaries \cite{1996A&A...313..497F,2000A&A...360..952H,2024A&A...682A.123T}. Convective overshooting allows material from convective zones to mix into adjacent radiative layers and vice versa. Overshooting of a convective zone undergoing nuclear burning increases the available nuclear fuel, leading to a greater mass of the ashes of the nuclear reactions. For example, overshooting causes a more massive helium core during the main-sequence phase of the star \cite{1997MNRAS.285..696S,2019A&A...622A..50H,2024A&A...682A.123T}. The larger helium core mass increases the rate at which this nuclide is burned when it ignites; as a result, the star has a higher luminosity and cooler effective temperature compared to a star where overshooting did not occur \cite{1997MNRAS.285..696S}. 
The same effect occurs for the carbon-oxygen (CO) core and subsequent carbon burning. 

Overshooting can alter the evolution of the star to the extent that a star that might be expected to yield a CO white dwarf without overshooting instead explodes as a core-collapse supernova \cite{2024A&A...682A.123T}. Stronger overshooting prescriptions are more likely to cause interactions between burning regions, affecting the pre-supernova structure \cite{2026MNRAS.546f2245W, 2019MNRAS.484.3921D}. Overshooting has another effect once deleptonization of the core begins at the initiation of core oxygen burning. Deleptonization occurs more quickly at the center of the star, establishing a gradient in the electron fraction of the stellar material. Convection moves material with higher electron fraction down into the core and thus alters the isotopic composition. This effect is particularly important during the core silicon burning phase in the last few days of the star's life, when the composition in the core approaches nuclear statistical equilibrium and the neutrino emission from beta processes grows significantly. Any changes in the composition of the material in the core will alter the beta neutrino emissivity, which thus depends upon the overshooting treatment adopted. The detection of pre-SN neutrino emission from an RSG would allow us to directly probe late burning in the core of the star and better constrain stellar evolution theory. 

The purpose of our study is to examine the sensitivity of pre-SN neutrino emission to the treatment of mass loss and convective overshooting. To do this, we construct a grid of models using the \MESA stellar evolution code \cite{Paxton_2011}. 
In \S\ref{sec:modeling} we describe our grid and the prescriptions and parameters we adopt for this study. In  \S\ref{sec:computingpreSN} we describe how the pre-SN neutrino spectra are computed. In \S\ref{sec:results1} we present the results of the stellar modeling, focusing first upon the surface properties and then upon the core of the stars. In \S\ref{sec:results2} we use the data from the models constructed in \S\ref{sec:results1} to compute the pre-SN neutrino emission. In \S\ref{sec:6} we explore the detectability of these differences in neutrino emission. In \S\ref{sec:conclusions} we summarize our findings and form our conclusions.      


\section{Modeling Red Supergiants with \MESA} 
\label{sec:modeling}

Evolving stars to collapse presents significant challenges.
Core structure and composition depend not only on the treatment of uncertainties in stellar evolution, but also on numerical and resolution choices, such as mass resolution \cite{2016ApJS..227...22F}, dimensionality \cite{Fields_2022, 2020MNRAS.491..972A}, and the size of the nuclear network \cite{2016ApJS..227...22F,2024RNAAS...8..152R}. The later-burning phases are computationally demanding, becoming ever more so with higher resolution and larger nuclear networks. Our highest priority when creating our RSG models was to include a sufficiently large nuclear network to converge on the core composition, which is necessary to compute the pre-SN neutrino emission. Models evolved with small networks, e.g., \texttt{approx21\_plus\_cr56}, are known to be unsuitable for supernova progenitor studies \cite{2024RNAAS...8..152R}, and the size of the nuclear network generally has a stronger effect on the pre-SN core structure than mass resolution and other numerical controls \cite{2016ApJS..227...22F}. Efforts have been made to emulate the inclusion of large nuclear networks to reduce computational burden, but currently stellar evolution codes incorporate the full networks and solve these equations with implicit methods \cite{2025ApJS..279...49G}. Consequently, we based our choices of numerical controls and tolerances around achieving the primary objective of successfully reaching core collapse with a large network.  
Following the study by Farmer \emph{et al.} \cite{2016ApJS..227...22F}, which found about 200 isotopes are needed to converge on the core composition, we employ the 206-isotope network \texttt{mesa\_206} from the main sequence to the onset of core collapse. The \texttt{mesa\_206} network was designed to reproduce results from a 1,200-isotope network with an error $\leq3\%$ (see \cite{1995ApJS..101..181W,1999ApJS..124..241T,FXT}). 
We use the large network throughout the evolution because this suppresses ``breathing pulses" \cite{1985ApJ...296..204C} during the earlier evolutionary phases \cite{2016ApJS..227...22F}; see also \cite{1989ApJ...340..241C,1989ApJS...71...47C} for other solutions on suppressing breathing pulses.

We use \MESA release 24.03.1 and evolve our models from the main sequence to the onset of core collapse, which we define as the time when iron core infall velocity exceeds 300 km/s. To be clear, as the models approach core collapse and nuclear statistical equilibrium, we maintain the use of the full nuclear network and do not switch to assuming nuclear statistical equilibrium. It is important to note that this release and all previous releases of \MESA are subject to a significant code error in the calculation of reverse reaction rates involving three or more products. This code error was shown to profoundly affect the equilibrium composition for large nuclear networks \cite{MESA_Issue_575}. We implemented a fix for this code error published on the \MESA GitHub repository \cite{MESA_PR_632} so our models would not be affected by this problem.
All of our models assume solar metallicity \cite{2009ARA&A..47..481A} and are nonrotating. To reduce the computation time and potentially improve numerical stability, we use operator split burning for $T\geq2.5 \times 10^9$ K \cite{mesa_vi}. We set the mixing length parameter $\alpha_{\rm MLT}=1.5$, to facilitate comparisons with many previous studies of RSGs (e.g., \cite{Heger_2001, 2016ApJS..227...22F, 2017ApJ...840...10Y, 2019A&A...625A.132S, 2020A&A...638A..55K, 2021MNRAS.505.4874H, 2021A&A...645A..54K, 2022A&A...662A..56K, 2022MNRAS.514.3736S, 2023A&A...678L...3V}) although we note that calibrations of $\alpha_{\rm MLT}$ from the comparison of models with observed effective temperatures $T_\mathrm{eff}$ favor larger values of $\alpha_{\mathrm{MLT}}$ (see, e.g., \cite{2018ApJ...853...79C, 2022ApJ...939...28C}) even though there are caveats about such comparisons; see Section \ref{sec:results1} for a discussion. We use \texttt{MLT\_option = TDC}, employing time-dependent convection based on \cite{1986A&A...160..116K} for the mixing length theory (MLT) \cite{1958ZA.....46..108B}. For the atmosphere we use the following options: \texttt{atm\_option = T\_tau}, \texttt{atm\_T\_tau\_opacity = fixed}, and \texttt{atm\_T\_tau\_relation = Eddington}. For details of these atmosphere implementations we refer the reader to the \texttt{MESA} atmosphere module documentation. We apply the Ledoux criterion for convective instability and set the semiconvection efficiency $\alpha_{\mathrm{SEM}}=0.01$, a choice that is common in previous studies of massive stars \cite{2016ApJS..227...22F, 2018ApJS..234...19F, 2018ApJ...853...79C} because inefficient semiconvection is consistent with small post-main-sequence YSG/BSG populations (see, e.g., \cite{2023A&A...680A.101S}). We use a set of zero-age main sequence masses \MZAMS of $\{ 12, 15, 18, 20\}$ \Msun. Numerical issues prevented us from considering higher or lower ZAMS masses.

Our model grid includes two overshooting prescriptions, which we call the ``core only" and ``all boundaries" treatments. These prescriptions are a mixture of step overshooting and exponential overshooting.  Step overshooting simply sets a constant diffusion coefficient $D(r)$ across the radiative-convective boundary nominally located at $r_0$, the radius where the radiative temperature gradient $\nabla_{\mathrm{rad}}$ is equal to the Ledoux gradient $\nabla_\mathrm{L}=\nabla_\mathrm{ad}+B$, where $\nabla_\mathrm{ad}$ is the adiabatic temperature gradient and $B$ is the composition (Brunt) term, and the fluid is marginally stable to convection. In chemically homogeneous regions, $B = 0$ and this reduces to the Schwarzschild criterion $\nabla_\mathrm{rad}=\nabla_\mathrm{ad}$. Since the diffusion coefficient vanishes at $r_0$, a representative value $D_{\rm conv}$ is taken at a radius slightly inside the convective region $r_{\rm conv} = r_0 - f_{0} H_{P,0}$, where $H_{P,0}$ is the value of the pressure scale height $H_P= -P / (dP/dr)$ at $r_0$, and $f_{0}$ is a parameter that sets the sampling depth. The sampled diffusion coefficient $D_{\rm conv}$ is applied over a distance $f\,H_P$ from this sampling point, i.e.,\ to a radius ($r_{\rm max}=r_0+ (f-f_0)\,H_P)$, with $f$ a parameter that controls the width of the convective overshooting layer. Exponential overshooting is similar to step overshooting, except the diffusion coefficient decreases exponentially as a function of distance starting from $r_{\rm conv}$ according to the equation
\begin{equation}
D(r) = D_{\rm conv} \exp \left( \frac{-2 (r - r_{\rm conv}) }{f H_{P,0}} \right)
\end{equation}
In both step and exponential overshooting, the temperature gradient is unchanged. For a thorough discussion of these overshooting schemes and their application in stellar evolution codes in general we refer the reader to \cite{2023Galax..11...56A}, and to  \cite{1996A&A...313..497F,2000A&A...360..952H} and the \MESA documentation for details about the implementation of overshooting in \MESA.

In our ``core only'' treatment of overshooting, we apply step overshooting across the hydrogen core with $f=0.345$ and $f_{0}=0.01$ \cite{Brott2011}, and exponential overshooting across the helium core with $f=0.01$ and $f_{0}=0.005$ \cite{2000A&A...360..952H, Paxton_2011, paxton_2013}. No overshooting is applied across any other radiative-convective boundary throughout the star during hydrogen and helium burning, and no overshooting is applied after core helium exhaustion. 
In the ``all boundaries" treatment, we follow the ``core only'' treatment across the hydrogen and helium cores and also apply weak exponential overshooting ($f=0.005$, $f_0=0.001$) \cite{Paxton_2011, paxton_2013} across all other radiative-convective boundaries throughout the star. After core helium exhaustion, weak exponential overshooting is applied across the core and all other radiative-convective boundaries until core collapse, except the carbon-oxygen core/helium shell boundary, which we avoid due to numerical difficulties. This ``all boundaries" overshooting treatment is constructed so as to investigate the effects of making a small change to the assumed overshooting prescription---that is, including weak overshooting at these other boundaries compared to none at all in the ``core only" treatment---rather than varying the strength of a single prescription. For more comprehensive studies of how varying the strength of the overshooting prescription affects the core structure, see \cite{2026MNRAS.tmp...19W} (exponential overshooting) and \cite{2024A&A...682A.123T} (step overshooting).

For the mass loss in our models, we adopt the so-called ``Dutch" scheme, employing the Vink \emph{et al.} \cite{Vinketal2001} prescription for hot phases and the de Jager, Nieuwenhuijzen, and van der Hucht \cite{1988A&AS...72..259D} prescription for cool phases. 
The mass lost by the star during the hot phase is negligible (but not ignored); the mass-loss rate during the cool phase is given by the empirical equation
\begin{equation}
     \log \left( \frac{-\dot{M}}{\eta} \right)  =  -8.158 +1.769 \log \left( L_{\star} \right)  -1.676 \log \left( T_{\mathrm{eff}} \right) 
\label{eq:masslossrate}
\end{equation}
In this equation, $\eta$ is a wind efficiency scaling parameter (also used during the hot phase to scale the mass loss rate), $L_{\star}$ is the surface luminosity of the star in units of the solar luminosity, and $T_{\text {eff }}$ is the effective surface temperature. 
The brighter the star and/or the lower its effective temperature, the higher the mass-loss rate. We vary the linear scaling parameter $\eta$, choosing values of 0.2, 0.4, 0.8, and 1.0. Modern simulations sometimes adopt values of $\eta<1.0$ to account for wind inhomogeneities (``clumping") and other evidence suggesting the standard ``Dutch" prescription with $\eta=1.0$ is too aggressive \cite{Puls2008,2014ARA&A..52..487S}. 
We only consider the ``Dutch'' scheme because Renzo et al. \cite{2017A&A...603A.118R} showed that the choice of scaling parameter $\eta$ is generally much more influential than the choice of mass loss prescription.
We disable mass loss when the core temperature satisfies $T_{\mathrm{center}}>1.1\times10^9$ K because the extreme mass-loss rates encountered in the late phases of the star's life reduce the size of the time steps so significantly that the computational cost to reach collapse becomes excessive. This temperature is reached approximately in the late stages of carbon burning, at which time the life expectancy of the star is only a few years. Turning off the mass loss at this time allows the model to reach collapse at lower computational expense, while being late enough that the amount of mass lost and change to the surface properties after this time are negligible.

In addition, per the recommendation in the \MESA source code, we applied an extra pressure factor of 2.0 at the surface to stabilize the stars' atmospheres because we frequently encountered problems in helium burning and later phases when super-Eddington zones caused instabilities. The instabilities often lead to time steps of order $10^{-6}$ s or smaller, which prevent the model from reaching core collapse. An extra pressure factor of 1.5 is recommended where possible in the \MESA massive star test suite \cite{Paxton_2011}\cite{paxton_2013}. Applying extra pressure to the surface suppresses the inflation and reddening caused by the envelope nearing the Eddington limit.


\section{Computing The Pre-Supernova Neutrinos}
\label{sec:computingpreSN}

Using the density, temperature, and compositional profiles of the RSGs, we calculate the neutrino emission spectra for all flavors ($\nu_e, \bar{\nu}_e, \nu_x, \, \mathrm{and} \, \bar{\nu}_x$) from beta processes and the pair annihilation process. The contribution of other thermal processes is minimal in the late stages of stellar evolution, so we omit them here \cite{2017ApJ...840....2P, 2017ApJ...848...48K}.  The spectra are calculated for each radial zone of the \MESA output, and then integrated over the volume of the star.  

\subsection{Beta Processes}

Beta processes produce electron flavor neutrinos and antineutrinos through decays of nuclei and capture of electrons and positrons.  The neutrino spectrum for a beta process is completely determined by the phase space of the electrons/positrons involved and can be written as 
\begin{eqnarray}\label{eq:betaSpect}
\phi_{EC,PC} & = & \frac{N\,E_{\nu}^2 \left( E_{\nu} - Q_{ij}\right)^2\, \Theta\left( E_{\nu} - Q_{ij} - m_e \right)}{1 + \exp\left[\left( E_{\nu} - Q_{ij} - \mu_e\right)/kT \right]},\\
\phi_{\beta} & = & \frac{N\,E_{\nu}^2 \left( Q_{ij} - E_{\nu} \right)^2\, \Theta\left( Q_{ij} - m_e - E_{\nu}\right)}{1 + \exp\left[\left( E_{\nu} - Q_{ij} + \mu_e\right)/kT \right]},
\end{eqnarray} 
where $EC/PC$ is electron/positron capture, and $\beta$ is a decay.  Here, $\mu_e$ is the electron chemical potential, including the electron rest mass $m_e$ so that $\mu_{e^-} = -\mu_{e^+}$.  $E_{\nu}$ is the energy of the emitted neutrino, and $N$ is a reaction-specific normalization factor.  The $Q$-value is defined as $Q_{ij} = M_p - M_d + E_i - E_j$, for a transition from a parent with mass $M_p$ and excitation energy $E_i$ to a daughter with mass $M_d$ and excitation energy $E_j$.

We use an effective $Q$-value method as described in \cite{2001PhRvC..64e5801L} and \cite{2017ApJ...840....2P}.  In this method, the value of $Q_{ij}$ is adjusted until the spectrum reproduces the average energy provided in calculated nuclear tables.  The tables report rates that are actually the sum of the rate for all possible transitions, from any $E_i$ to any $E_j$.  Rather than calculating a particular $Q_{ij}$ for each possible reaction, we treat the process as though all of the strength is in a single transition.  Because the average energy in the tables is the average over both decay and capture, this method results in the same $Q$-value for both processes.  The rate tables we use are a combination from \cite{2001PhRvC..64e5801L} (LMP), \cite{ODA1994231} (OEA), and \cite{1980ApJS...42..447F, 1982ApJ...252..715F, 1982ApJS...48..279F, 1985ApJ...293....1F} (FFN). These are the same rates used by \MESA, and we follow the same order of precedence for isotopes found in multiple tables: rates from LMP are preferred, followed by OEA and then FFN.
The normalization constant $N$ is found by matching our effective $Q$-value spectrum to the rate given in the tables:
\begin{equation}\label{eq:betaRate}
\lambda_i = \int_0^{\infty} \phi_i \, dE_{\nu}, \,\,\,\, i \in \left\{ \mathrm{EC,\, PC,\, \beta^{\pm} } \right\}
\end{equation}
where $\lambda_i$ is the rate of the specific process for a single isotope. 

The complete spectrum is found by performing a weighted sum over all isotopes present as
\begin{equation}\label{eq:totalBetaSpect}
\left(\frac{dR_{\beta}}{dE_{\nu}}\right)_{\nu_e,\bar{\nu}_e} = \sum_k \frac{\rho}{m_p A_k}\,X_k\,\phi_k  , 
\end{equation}
where $X_k$ is the isotopic abundance of isotope $k$, $\rho$ is the mass density of the radial zone, $m_p$ is the mass of a proton, and $A_k$ is the mass number of isotope $k$. Note that this process can only create electron flavor neutrinos and antineutrinos due to flavor conservation.

We should also note here that the effective $Q$-value method is inherently limited due to its nature as a single strength approximation \cite{2001PhRvC..64e5801L, dzhioev_neutrino_2023}.  Dzhioev \textit{et al.} have calculated the neutrino energy spectra for a limited number of isotopes in pre-SN stars using a thermal quasiparticle random-phase approximation (TQRPA) technique, which allows for transitions to and from multiple thermally excited states \cite{dzhioev_neutrinos_2023, dzhioev_neutrino_2023, dzhioev_neutrinos_2024, dzhioev_pre-supernova_2025}.  These studies show that the effective $Q$-value method misses potentially important features of the spectrum, such as double-peaked structures and extended high energy tails.  We continue with the effective $Q$-value method to remain consistent with the rates and energies used in \MESA itself, as well as to allow efficient implementation for a large nuclear network.  

\subsection{Pair Annihilation}

Neutrinos can also be produced from the annihilation of electron-positron pairs.  Unlike beta processes, pair annihilation can produce neutrinos and antineutrinos of all flavors.  The differential rate of emission, as given in \cite{misiaszek_neutrino_2006}, is 
\begin{equation}\label{eq:pairSpect}
\left( \frac{dR}{dE}\right)_{\nu_{\alpha},\bar{\nu}_{\alpha}} = \int d^3 p_1 d^3 p_2 \left( \frac{v\,d\sigma}{dE}\right)_{\nu_{\alpha},\bar{\nu}_{\alpha}}  f_1\,f_2, 
\end{equation}
where $f_{1,2}$ are the Fermi-Dirac distributions for the electron and positron, and the cross section times the relative velocity of the pair $v$ is
\begin{equation}\label{eq:vdsigmaPair}
v\,d\sigma  = \frac{1}{16\,\pi^2\,\mathcal{E}_1\,\mathcal{E}_2} \delta^4(P_1 + P_2 - Q_1 - Q_2) \frac{d^3 q_1}{2 E_1} \frac{d^3 q_2}{2 E_2} \langle | \mathcal{M}|^2 \rangle.
\end{equation}
In this expression, $P_{1,2} = (\mathcal{E}_{1,2}, {\bf p}_{1,2} )$ is the four-momentum of the electron or positron, $Q_{1,2} = (E_{1,2},{\bf q}_{1,2})$ is the four-momentum of the neutrino or antineutrino, and $\langle | \mathcal{M}|^2 \rangle$ is the squared matrix element for the process.  The squared matrix element is found to be \cite{misiaszek_neutrino_2006}
\begin{widetext}
\begin{equation}\label{eq:pairM}
\langle | \mathcal{M}|^2 \rangle  = 8\,G_F^2  \left( \left( C_A^f - C_V^f\right)^2 \left(P_1 \cdot Q_1\right) \left( P_2 \cdot Q_2\right) + \left( C_A^f + C_V^f\right)^2 \left(P_1 \cdot Q_2\right) \left( P_2 \cdot Q_1\right) + m_e^2 \left( (C_A^f)^2 - (C_V^f)^2\right) \left(Q_1 \cdot Q_2\right) \right),
\end{equation}
\end{widetext}
where $m_e$ is the mass of the electron, and $f = e$ for $\nu_e$ or  $f=x$ for $\nu_{\mu,\tau}$. The coupling constants are
\begin{eqnarray}\label{eq:CvCa}
C_V^e & = &\frac{1}{2} + 2 \sin^2{(\theta_w)}\\
C_A^e & = &\frac{1}{2}\\
C_V^x & = & 1 - C_V^e \\
C_A^x & = & 1 - C_A^e.
\end{eqnarray}
The integration is done over all possible electron and positron momenta, using a Monte Carlo integration technique.  Specifically, the Cuhre algorithm from the \texttt{PyCuba} package is used to implement the Monte Carlo integration \cite{2014A&A...564A.125B}.


\section{Results}
\label{sec:results1}

\subsection{Pre-Collapse Observables}

Using the inputs described above, we successfully evolved 32 RSG models with \MESA to the point of core collapse, defined to be when the maximum infall velocity of a mass zone within the iron-group-rich core reaches 300 km/s. We begin summarizing the model results by focusing on the observable properties of the star: mass, luminosity, and effective temperature. 

\begin{figure}
    \centering
    \includegraphics[width=1.0\linewidth]{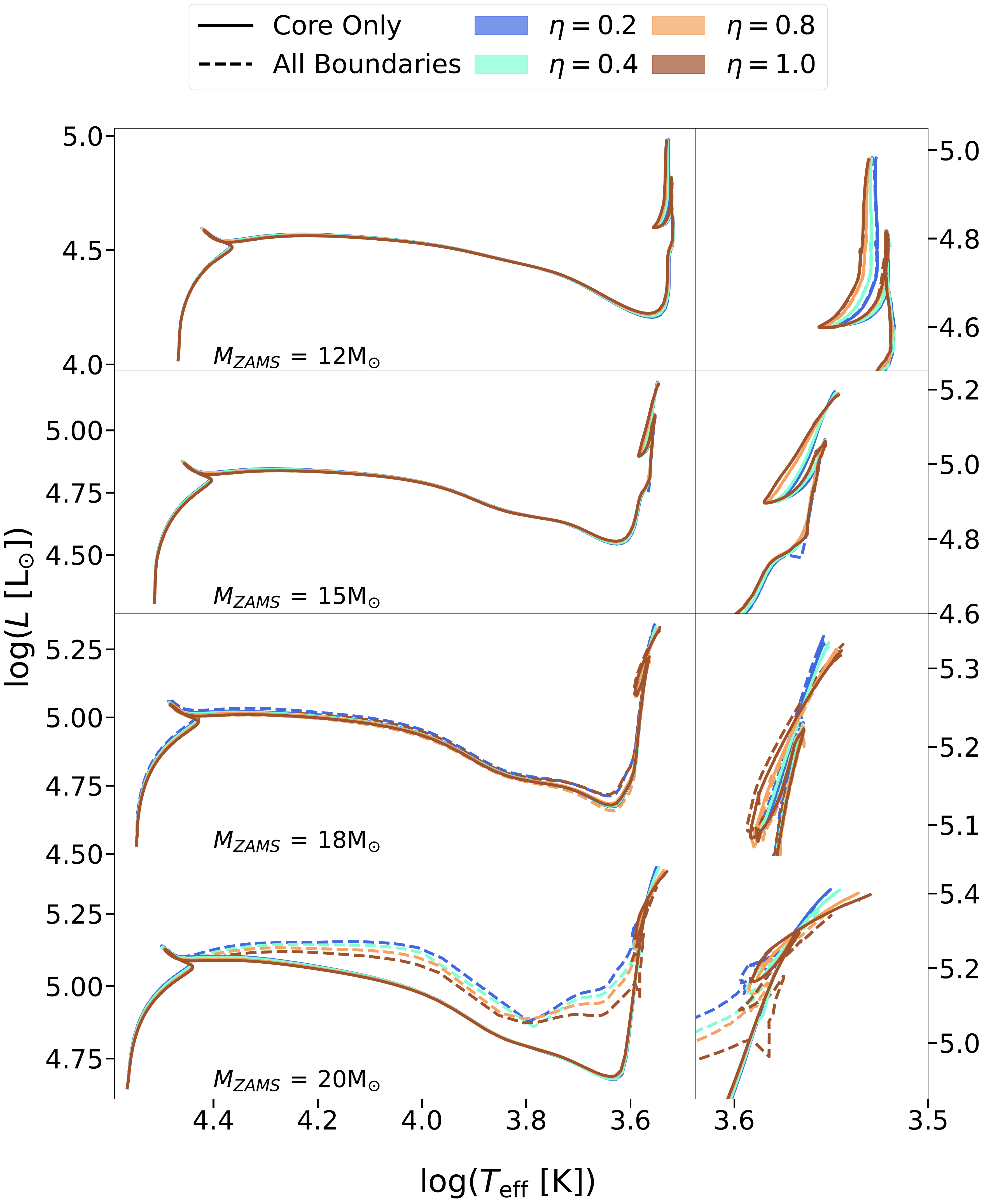}    
    \caption{The evolution of the luminosity and effective surface temperature (HR diagrams) of the model RSGs. }
    \label{fig:HR}
\end{figure}

Figure (\ref{fig:HR}) shows the evolutionary tracks of the stars in the luminosity-effective temperature plane. The trajectories begin when the stars reach the main sequence (at the left-hand side of each panel). After the main sequence, the stars move rightwards to cooler effective temperatures and larger radii without a significant change in luminosity. Then helium burning ignites, and it is during the helium burning phase that stars lose the most mass. In the small panels on the right-hand side, we have zoomed in upon the end of the trajectory for the reader to see in greater detail how the models differ during the final phases. The figure shows that the HR tracks for stars with $\MZAMS< 18$ \Msun are largely independent of mass loss and overshooting treatment until the star reaches the very last stage of its life. While higher mass-loss efficiency corresponds to higher surface temperature over the majority of a star's evolution, we see that for $\MZAMS\geq 18$ \Msun, the trajectories for the different mass-loss efficiencies cross so that higher mass-loss efficiencies result in cooler pre-SN surface temperatures in the period leading up to collapse. This crossing occurs because of two competing effects. Stars with higher mass-loss efficiencies in general have smaller radii during helium burning, causing them to have higher $T_\mathrm{eff}$. However, after core helium exhaustion, the stars expand and brighten, and stars retaining a smaller fraction of their original mass in the envelope expand and cool more (consistent with \cite{snapshot}). Stars with higher \MZAMS are more luminous and thus lose more of their envelopes (see equation \eqref{eq:masslossrate}).

\begin{figure}
    \centering
    \includegraphics[width=1.0\linewidth]{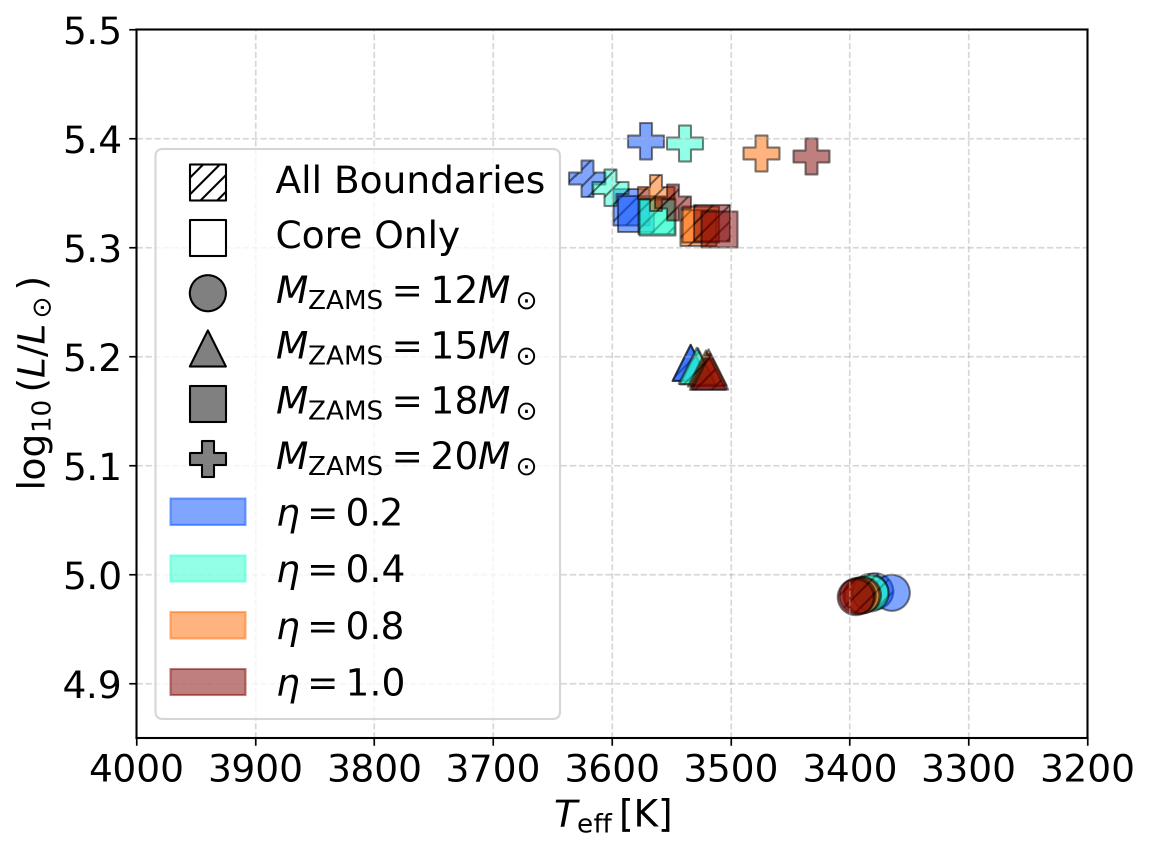}    
    \caption{The luminosity and effective surface temperature of the model RSGs at the point of collapse.}
    \label{fig:endHR}
\end{figure}

Figure (\ref{fig:endHR}) shows the positions of the model stars on the HR diagram 10 minutes pre-collapse. As is evident in figure (\ref{fig:HR}), for $\MZAMS\leq18\;\Msun$, the mass loss and convective overshooting prescriptions only weakly affect the pre-SN luminosity of the star. For $\MZAMS \leq 18 \;\Msun$, the pre-SN effective temperature is also only weakly affected by mass loss and the details of the overshooting. Only above this $\MZAMS$ do the final positions of the models on the HR diagram begin to separate. For $\MZAMS = 20 \;\Msun$, we observe that stars with stronger mass loss are slightly less luminous and cooler pre-SN than their counterparts with weaker mass loss. For $\MZAMS = 20\;\Msun$, the ``core only'' models are more luminous than ``all-bounds'' models. The reason the luminosities do not appreciably differ between models of the same ZAMS mass below $\MZAMS=\:20\:\Msun$ is that the surface luminosity depends largely on the helium core mass \cite{snapshot}, and helium core mass is barely affected by the mass loss and overshooting treatments for lower ZAMS masses. This is expected because the overshooting across the hydrogen and helium core are identical in the two overshooting treatments, and mass loss only directly affects the envelope; its influence on the helium core is mediated by mass loss on the main sequence---only a small amount of the total mass lost over the star's life---reducing pressure on the core, and by mixing. In the ``all boundaries'' $\MZAMS=20\:\Msun$ models, overshooting at the base of a convective zone that forms above the hydrogen shell allows the zone to extend close to the helium core, mixing the helium-enriched material outward and more hydrogen-rich material inward toward the edge of the core. Therefore, more material must be burned to grow the helium core, so the core grows more slowly (see \cite{2020MNRAS.496.1967K, 2023A&A...680A.101S}), leading to lower luminosity than the ``core only'' models. In the ``core only'' models, the lack of overshooting in this region prevents this interaction. Meanwhile, $T_\mathrm{eff}$ depends on the radius of the star, which is affected by mass loss. The effective surface temperatures ($T_{\mathrm{eff}}$) of our models shown in figure (\ref{fig:endHR}) can be compared with observations, such as those by \cite{2005ApJ...628..973L,2006ApJ...645.1102L,Davies_2008,Davies_2013}, and other model grids \cite{2012A&A...537A.146E,2013A&A...558A.131G}. However, we caution the reader that (a) the $T_{\mathrm{eff}}$ of RSGs inferred from observations depend upon the method used \cite{Davies_2013,2019AJ....157..167D}; (b) the $T_{\mathrm{eff}}$ reported by \texttt{MESA} is defined via a model atmosphere controlled by parameters we have not investigated in this study, and which is also different---and simpler---than the ones used to infer $T_{\mathrm{eff}}$ from observations;  and (c) the $T_{\mathrm{eff}}$ shown in figure (\ref{fig:endHR}) are taken immediately before collapse, whereas the observations will be for a stage prior to this time. As shown in figure (\ref{fig:HR}), the $T_{\mathrm{eff}}$ of the models are decreasing prior to collapse; therefore, the model $T_{\mathrm{eff}}$ shown in figure (\ref{fig:endHR}) should be systematically lower than observations, all the other caveats aside. Keeping in mind these cautionary statements, the $T_{\mathrm{eff}}$ of our models are seen to lie between $3350\;{\rm K} \lesssim T_{\mathrm{eff}} \lesssim 3650\;{\rm K}$, which overlaps well with the observed temperatures by \cite{Davies_2008}---see also \cite{2017RSPTA.37560270D}---but less so with those by \cite{2005ApJ...628..973L}. The luminosities of our models are a good match with observations. 
The effective temperatures of the models by \cite{2012A&A...537A.146E,2013A&A...558A.131G} for the same 
mass range lie between $3500\;{\rm K} \lesssim T_{\mathrm{eff}} \lesssim 3700\;{\rm K}$.

\begin{figure}
    \centering
    \includegraphics[width=1.0\linewidth]{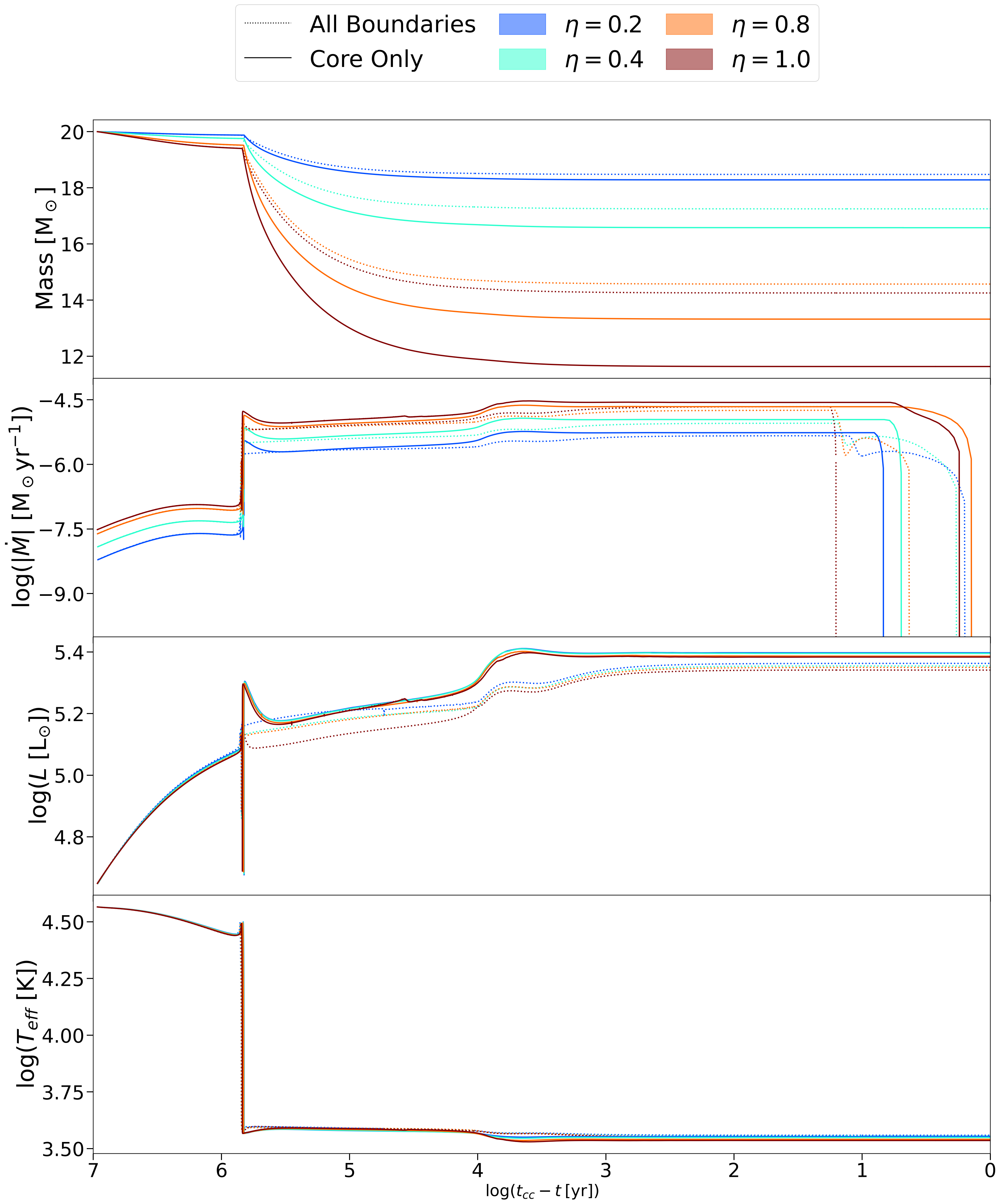}
    \caption{The evolution of the mass, mass loss rate, luminosity, and effective surface temperature of the $M_{\mathrm{ZAMS}}=20.0$ models as a function of time before collapse.}
    \label{fig:LTeffMdot}
\end{figure}

The luminosity and effective temperature control the mass loss rate, as shown in equation (\ref{eq:masslossrate}). In figure (\ref{fig:LTeffMdot}) we plot these quantities as a function of time for the 20 \Msun models together with the resulting mass loss rate and mass. We see that the mass loss rates jump upwards after the models cross the HR gap and increase again when helium shell burning is initiated, causing the outer layers of the star to further inflate. While the effective temperatures are very similar in all models, the luminosities of the ``core only" models are all larger than the corresponding ``all boundaries" models. As a result, mass loss rates are always larger for the ``core only" models. This is found to be a general result. As the top panel of the figure shows, most of the mass lost occurs during core helium burning and the difference in the final mass of the star increases with $\eta$. For $\eta = 1.0$, the difference in the final mass can be as large as $\approx 2$ \Msun. So while the mass loss rates during the later stages of the star's life can be larger than during core helium burning, the much shorter duration of those stages means that the mass of the star is not greatly changed over them. This justifies our protocol to turn off mass loss in the late stages of carbon burning. 

\begin{figure}
    \centering
    \includegraphics[width=1.0\linewidth]{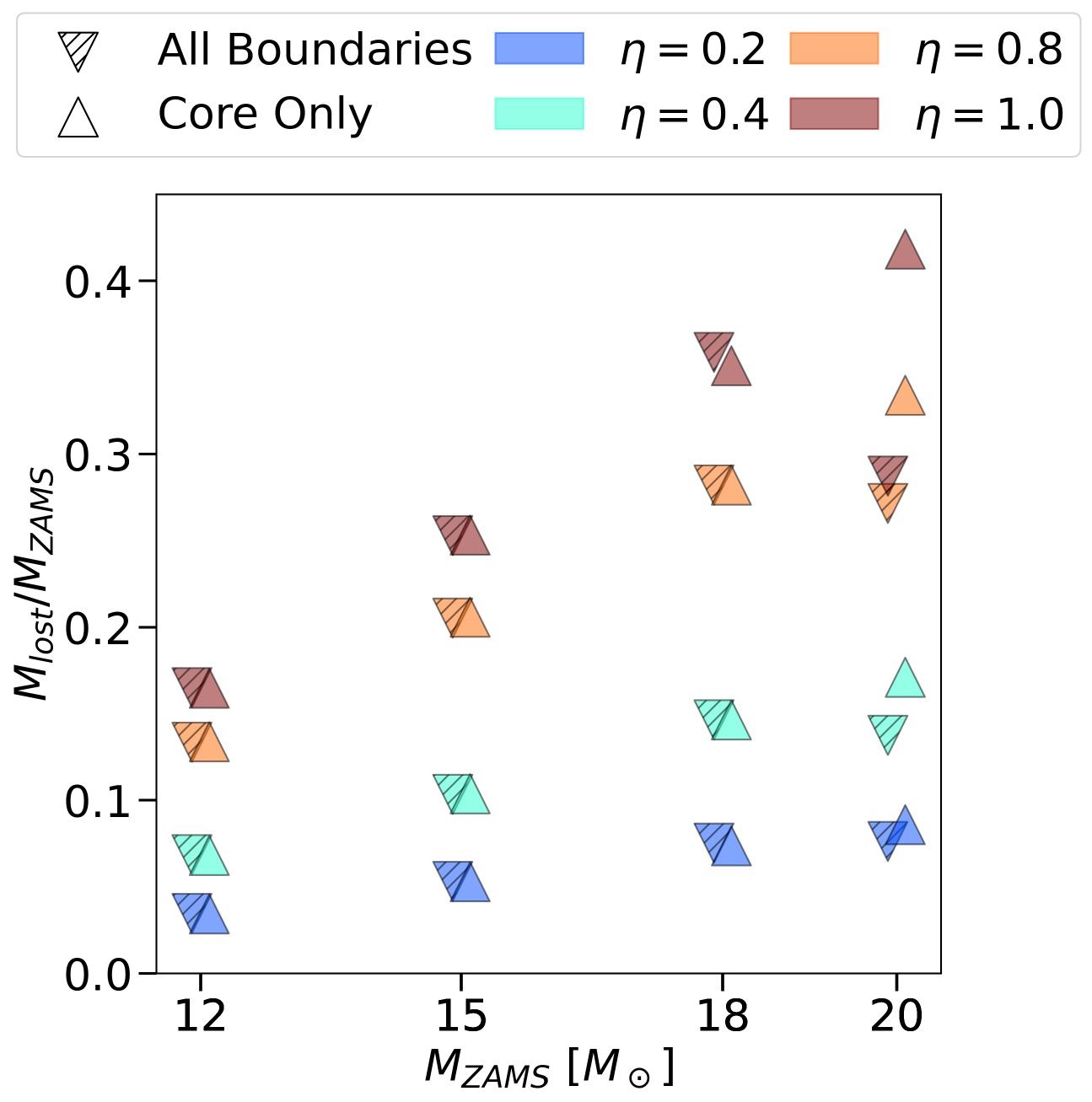}    
    \caption{The fraction of initial mass lost at the point of collapse for each of the RSG models. For readability ``all boundaries'' models are offset slightly to the left, and ``core only'' models are offset slightly to the right.}
    \label{fig:mass_lost}
\end{figure}

The fraction of the initial mass lost over the lifetime of the star is shown in figure (\ref{fig:mass_lost}). Despite the inherent non-linear response of the models to the input parameters, higher mass-loss efficiencies $\eta$ consistently lead to a greater fraction of the initial mass lost. For $\MZAMS\leq18\;\Msun$, we find almost no difference in total mass lost between the two overshooting prescriptions, but for $\MZAMS = 20\;\Msun$, the ``all boundaries'' models lose less mass than ``core only'' models. This is because, as shown in equation (\ref{eq:masslossrate}), the mass loss rate depends upon the luminosity, and for $\MZAMS=20\;\Msun$ the luminosities of the models begin to differ substantially between the overshooting treatments, as seen from the HR tracks in figure (\ref{fig:HR})). This $\MZAMS$ turnoff is also seen in \cite{2024A&A...682A.123T}, where the authors studied the effects of varying the strength of the step overshooting parameter $f$ (denoted $\alpha_{ov}$ in their work). They find that models with higher $f$ lose less mass than models with lower $f$ for $\MZAMS\geq21\;\Msun$. These authors note that for $\MZAMS<21\;\Msun$, the trend appears to be the opposite, but in our models we do not see a difference between overshooting treatments for $\MZAMS\leq$ $18\;\Msun$ because the luminosities are nearly indistinguishable for much of the HR track in figure (\ref{fig:HR}).
If we limit ourselves to the ``core only'' overshooting prescription, the fraction of the initial mass lost increases monotonically with \MZAMS; this is not the case for the ``all boundaries'' models, where the $\MZAMS = 20\;\Msun$ models lose a lower fraction of their initial mass compared to $\MZAMS = 18\;\Msun$.
Our results shown in figures (\ref{fig:HR}) through (\ref{fig:mass_lost}) mirror those of \cite{2017A&A...603A.118R}, \cite{2024A&A...682A.123T}, and \cite{2026MNRAS.tmp...19W}. 


\subsection{Core Properties: Temperature, Density, and Compactness}

\begin{figure}
    \centering
    \includegraphics[width=\linewidth]{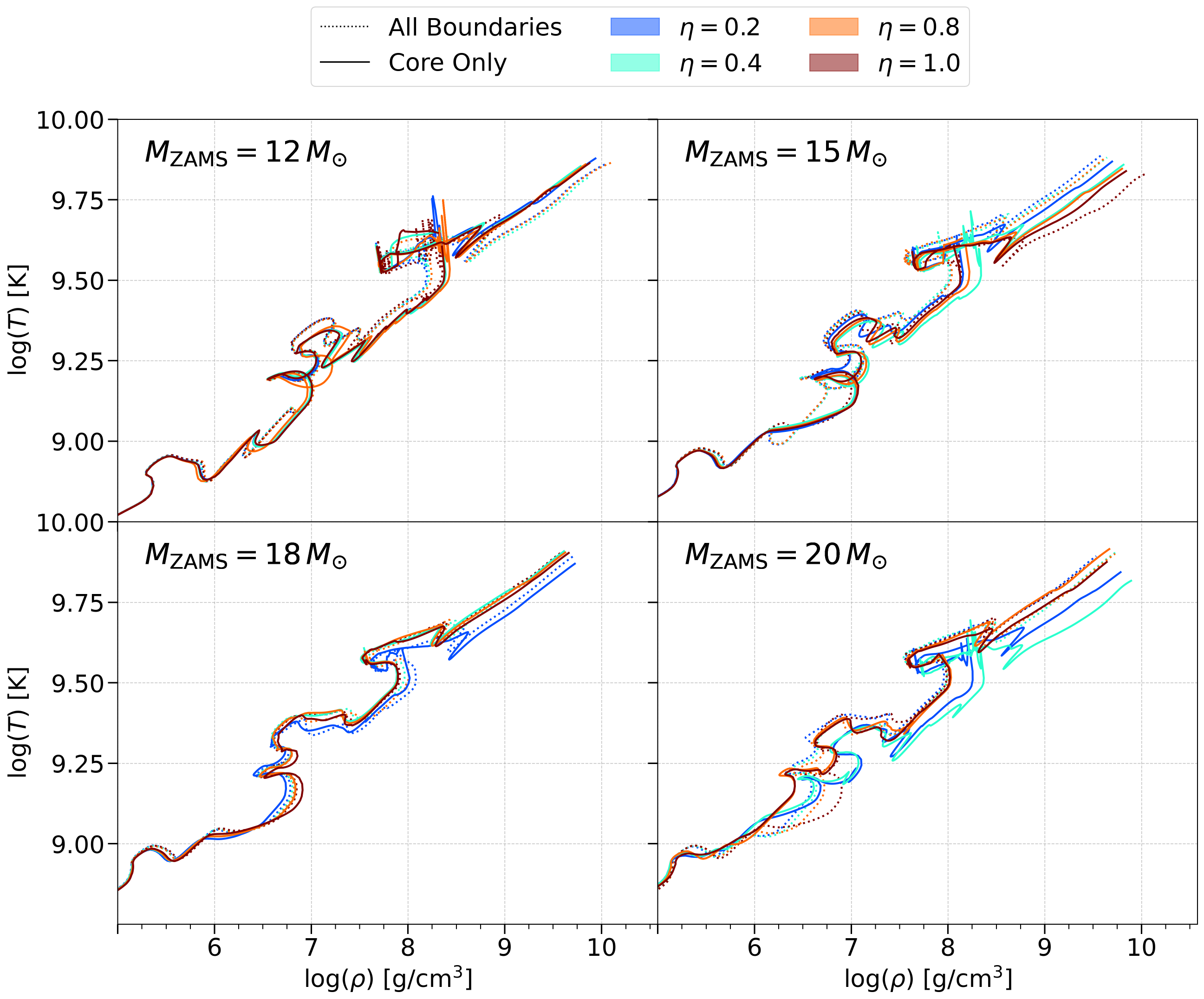}    
    \caption{The evolution of the central temperature and density for all models.}
    \label{fig:logrhologT}
\end{figure}

We now turn our attention to the core of the stars from which the neutrinos are emitted. In figure (\ref{fig:logrhologT}) we show the evolution of the central temperature and density for every model. The differences between the mass-loss efficiencies and overshooting treatment are now apparent even for \MZAMS = 12 \Msun. The trajectories begin to diverge starting around carbon burning, i.e.,\ when the central temperature and density exceed $\sim 5 \times 10^{8}\;{\rm K}$ and $\sim 3 \times 10^{6}\;{\rm g/cm^3}$. As the core becomes denser and hotter as the star approaches collapse, the trajectory becomes noticeably ``wiggly" \cite{RevModPhys.74.1015}, especially so at the point of core silicon ignition, when the central temperature and density exceed $\sim 3 \times 10^{9} \approx 10^{9.5}\;{\rm K}$ and $\sim 10^{8}\;{\rm g/cm^3}$ \cite{RevModPhys.74.1015} \cite{Iliadis}. These ``wiggles'' are due to the ignition and extinguishing of nuclear burning episodes \cite{RevModPhys.74.1015}. As equation (\ref{eq:totalBetaSpect}) indicates, the beta neutrino emission is proportional to the density, and the thermal neutrinos are also functions of $\rho$ and $T$ via the distributions of the electron-positron momenta in equation (\ref{eq:pairSpect}). Thus, the differences in the mass loss and overshooting treatment will affect the pre-SN neutrino emission starting with carbon burning, and the differences grow as collapse is approached. 

\begin{figure}
    \centering
    \includegraphics[width=1.0\linewidth]{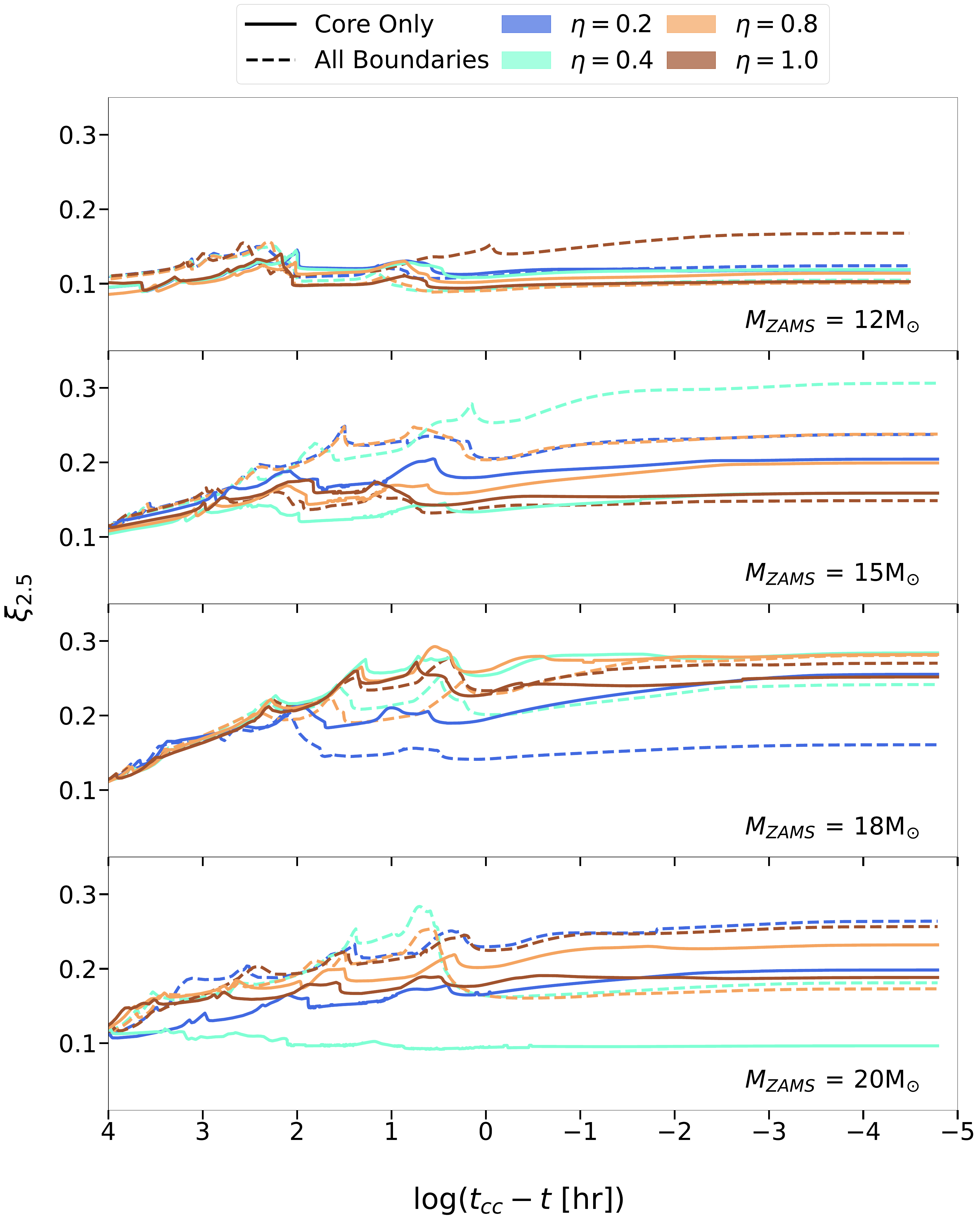}    
    \caption{The compactness $\xi_{2.5}$ as a function of time before collapse for all models.}
    \label{fig:compactnessvstime}
\end{figure}

The central density is a useful diagnostic of the differences between models, but a related quantity, called the ``compactness", is seen to better describe the entire core region. Following \cite{2011ApJ...730...70O}, the compactness of the star is defined to be
\begin{equation}
    \label{eq:compactness}
    \xi_M = \frac{M/\Msun}{R(M)/1000\;\mathrm{km}},
\end{equation}
where $M$ is a chosen mass and $R(M)$ is the radius (in km) that encloses that amount of mass. The compactness of the core of the RSG at ``bounce" has been shown to be a good, but not perfect, predictor of the final outcome of the eventual supernova \cite{2011ApJ...730...70O,2025arXiv250106784B}, and furthermore, the compactness of the core can also be determined from the subsequent supernova neutrino burst signal \cite{2017JPhG...44k4001H,2021arXiv210110624S}. Figure (\ref{fig:compactnessvstime}) shows the compactness $\xi_{2.5}$ of our RSG models as a function of the time before collapse. 
In many cases the compactness barely changes over the last 10,000 hours ($10,000\;{\rm h} \approx 1\;{\rm yr}$) of the star's life. However, there are many other models where the compactness is seen to gradually rise over time, but the rise is not monotonic; there are times when the compactness decreases. 
Similar to \cite{2025A&A...695A..71L}, we find that these decreases in compactness are correlated with episodes of nuclear burning, either in the core or a shell. We also point out the $\MZAMS = 18\;\Msun, \eta=0.2$, ``all boundaries'' model as an exceptional and interesting case. 
There is no obvious relationship between the compactness as a function of time and the mass-loss efficiency parameter or the overshooting treatment, but it is clear that the dispersion of the compactness evolution grows with \MZAMS. Davis et al. similarly found that for a $25\Msun$ star, varying the exponential overshooting parameter $f$ resulted in a large spread in $\xi_{2.5}$ \cite{2019MNRAS.484.3921D}.

\begin{figure}
    \centering
    \includegraphics[width=1.0\linewidth]{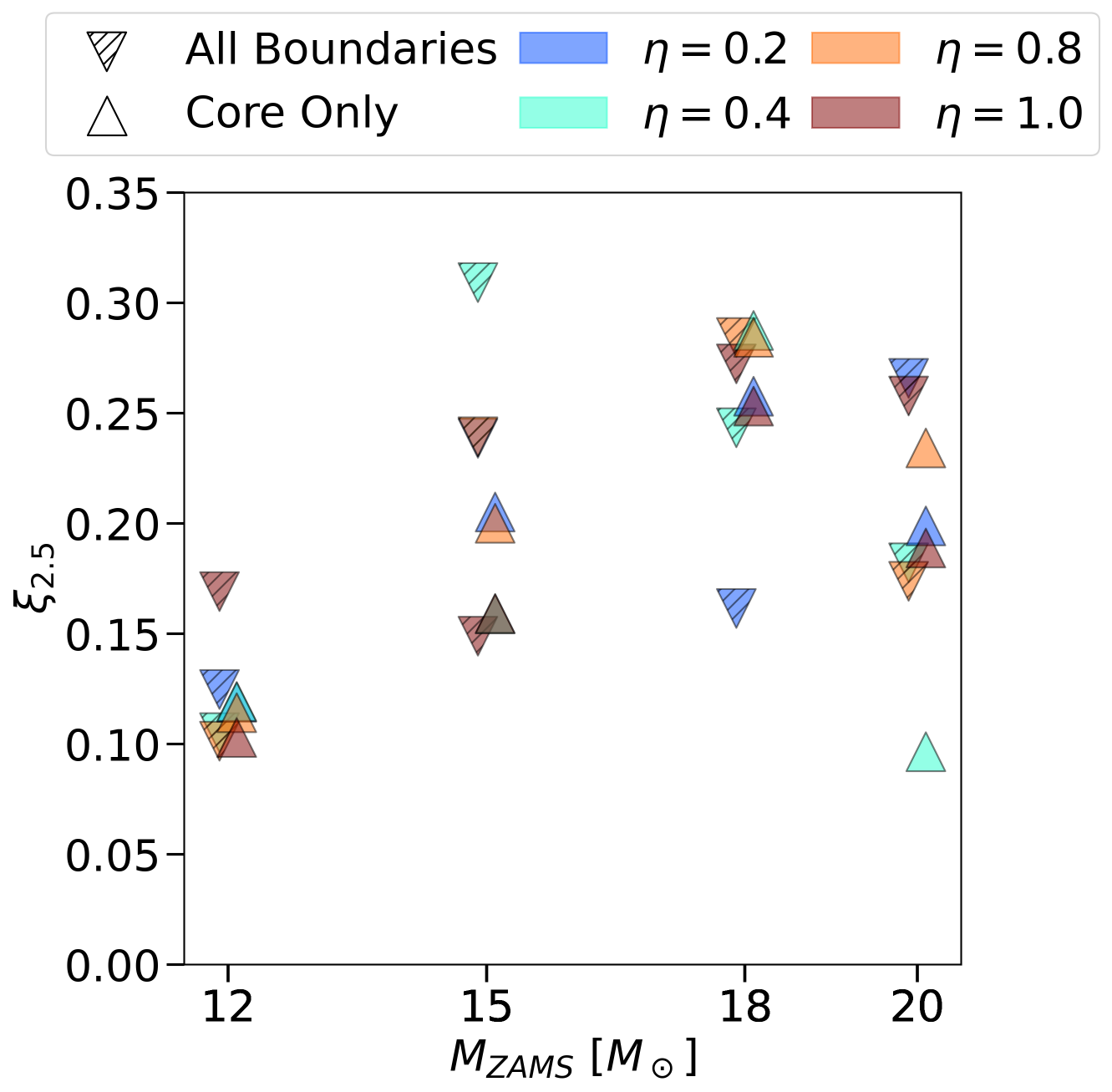}    
    \caption{The compactness $\xi_{2.5}$ at the point of collapse for all models.}
    \label{fig:compactnessatcollapse}
\end{figure}

In figure (\ref{fig:compactnessatcollapse}) we show the compactness at the point of collapse, defined to be the time at which the maximum infall velocity is $300\;{\rm km/s}$. Note that this time is before the time at which the compactness in supernova simulations is usually measured at ``bounce". This difference in the evaluation time means the compactness we calculate is slightly smaller than the compactness as measured at bounce, but, by choosing to use the $M = 2.5\;\Msun$ mass coordinate, the difference is very small \cite{2014ApJ...783...10S}. Figure (\ref{fig:compactnessatcollapse}) again shows that the uncertainty in the compactness of the model for a given ZAMS mass does depend upon the mass-loss efficiency and overshooting treatment, and that the dispersion grows with \MZAMS.


\begin{figure}
    \centering
    \includegraphics[width=1.0\linewidth]{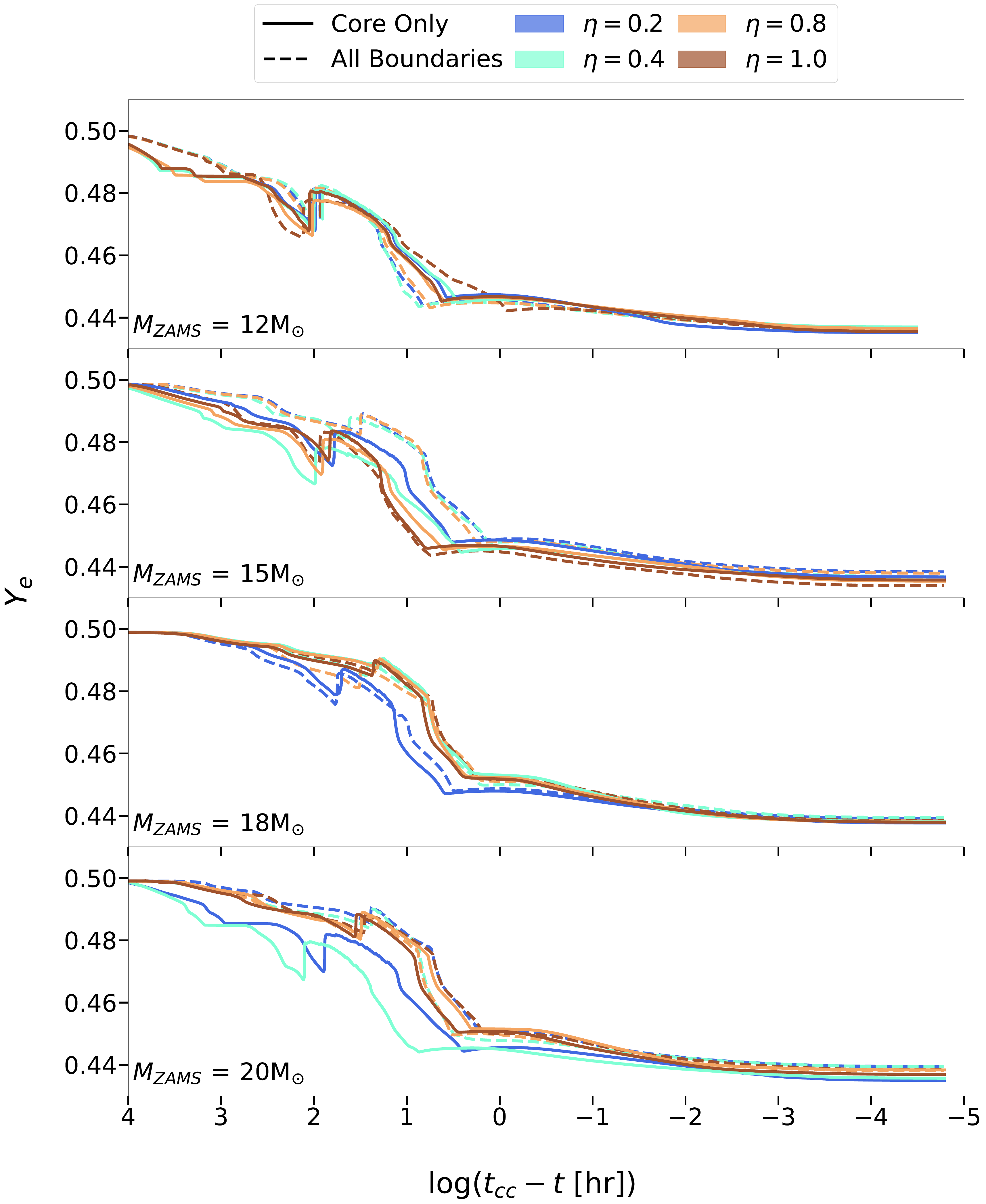}    
    \caption{The core electron fraction for all models as a function of the time before collapse.}
    \label{fig:coreYe}
\end{figure}

\subsection{The Core Composition}

In addition to the density and temperature, the other factor which influences the pre-SN neutrino emission is the composition, as seen in equation (\ref{eq:totalBetaSpect}). Since the star is so hot and dense as collapse is approached, the composition in the core region is very close to nuclear statistical equilibrium, which is determined solely by the density, temperature, and electron fraction $Y_e$. Note that in addition to its effect upon the composition, the electron fraction also influences the electron pressure and thus the structure of the core region, as well as influencing the maximum mass of the matter that can be supported by electron degeneracy pressure and thus the moment of collapse. 
In figure (\ref{fig:coreYe}) we show the electron fraction at the core of the stars as a function of time before collapse. This figure should be compared with figure (7) from \cite{2025A&A...693A..93G}. As late as $\approx 1$ year before collapse the core electron fraction is steady at the value of $Y_e = 0.4986$.  
Over the last year of an RSG's life, starting at core oxygen burning, the electron fraction begins to decrease as the electron captures plus nuclear beta decays begin to remove lepton number from the core. 
In the final hour of the star's life, the electron fraction of every model converges to $Y_e \approx 0.438$. Note that this final central electron fraction is substantially lower than seen in \cite{2025A&A...693A..93G}. Superposed on this general trend of deleptonization, we see that at some point between approximately 100 and 10 hours prior to collapse, there is a brief period when the electron fraction suddenly increases. Exactly when this increase in central electron fraction occurs, and how long it lasts, are highly variable and dependent on the mass loss and overshooting scheme.

\begin{figure}
    \centering
    \includegraphics[width=0.85\linewidth]{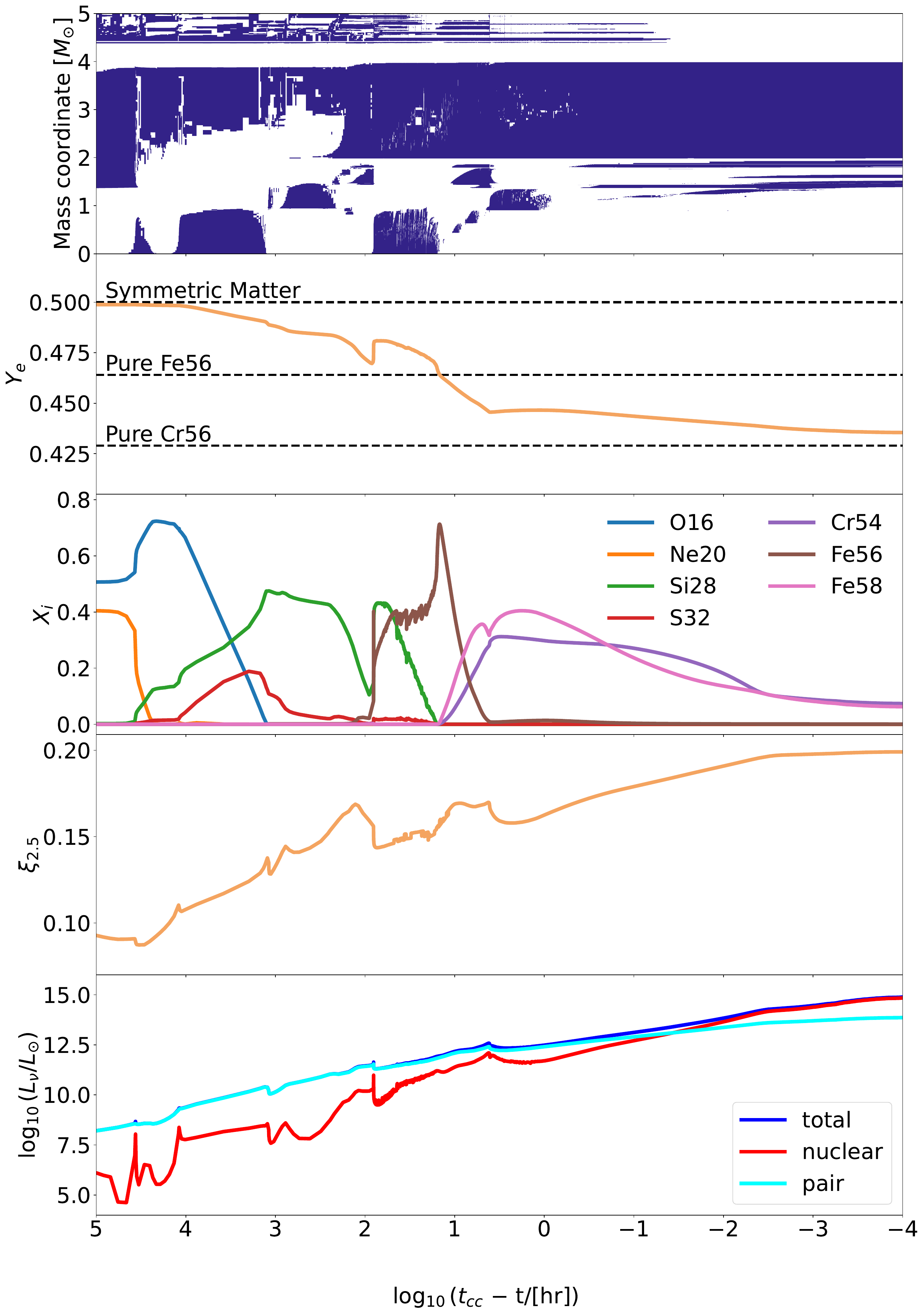}    
    \caption{The convective regions (top panel), core electron fraction (second panel), mass fractions (third panel), compactness (fourth panel), and neutrino luminosity (bottom panel) as a function of time before collapse for the 15 \Msun, $\eta = 0.8$ and ``core only'' overshooting model. In the top panel, convective regions are shown in blue, and radiative regions are shown in white.}
    \label{fig:yeconv1508}
\end{figure}   

To explain the origin of this feature, we plot in figure (\ref{fig:yeconv1508}) the regions of the star which are convective as a function of time before collapse for the 15 \Msun, $\eta = 0.8$, ``core only'' model. We see that the core becomes convective at three times in the last $10^5$ hours (approx 10 years) of the star's life: the first about two years before collapse is due to neon burning, the second episode about a year before collapse is due to oxygen burning, and the third about one hundred hours before collapse is due to silicon burning. These three episodes of convection in the core are seen elsewhere in the literature, e.g., \cite{2014ApJ...783...10S,2020ApJ...890...43C,2025A&A...693A..93G}, and the burning process is identified by the concomitant decrease in the mass fraction of the burning isotope as shown in the third panel. 
As noted previously, beginning with oxygen burning the core begins to deleptonize, and we have found that the principal reactions that deleptonize the core during this period are electron captures on aluminum-26, sulfur-33, chlorine-35, phosphorus-30, and argon-37. The deleptonization occurs more rapidly in the very center of the star so that the central electron fraction there is lower than in overlying material. Thus, when silicon burning begins and the core becomes convective again, the convection will mix material with higher electron fraction into the core, raising the electron fraction at the center of the star. This infusion of higher $Y_e$ material can be seen in the third panel where the silicon mass fraction at the center of the star suddenly  increases, and is thus the origin of the increase in the electron fraction seen in figure (\ref{fig:coreYe}). Prior to this enrichment of silicon, the mass fraction of silicon-28 had been decreasing since the end of core oxygen burning. At the end of core silicon burning, the material at the center of the star is predominantly iron-56 (in this model), but as the third panel shows, this quickly changes to more neutron-rich nuclei due to beta processes (mostly electron captures). 

The onset of a nuclear burning episode---either in the core or in a shell---not only changes the electron fraction, but also impacts the structure of the star, as shown in the fourth panel of figure (\ref{fig:yeconv1508}), where we plot the compactness $\xi_{2.5}$ of the star as a function of time. Starting with core neon burning, the compactness rises all the way through to the beginning of core silicon burning, whereupon it suddenly drops. The drop in compactness persists through core silicon burning, after which it recovers until another drop in compactness about 1 hour before collapse due to silicon shell burning, as seen in the top panel of the figure. See figure (4) in \cite{2020ApJ...890...43C} for similar changes in compactness due to shell burning episodes. Even after the silicon shell burning ends, the compactness continues to evolve significantly all the way until collapse due to neutrino losses, changing by $\sim 20\%$ over just the last hour of this star's life. The evolution of the compactness in this and the other RSG models we have constructed is similar to that seen by \cite{2014ApJ...783...10S,2020ApJ...890...43C}.

The evolution of the star shown in figure (\ref{fig:yeconv1508}) is typical of all the other models we have constructed, although the details change. The general trend is that the core begins to deleptonize starting with oxygen burning, reaches the electron fraction of iron-56 at the end of core silicon burning, and continues to decline thereafter due to electron captures. Although the final core electron fraction is almost identical for all models independent of the initial mass, mass-loss efficiency, and overshooting treatment, exactly how this value is reached depends upon the details and will be reflected in the neutrino emission.

\begin{figure*}
    \centering
    \includegraphics[width=\linewidth]{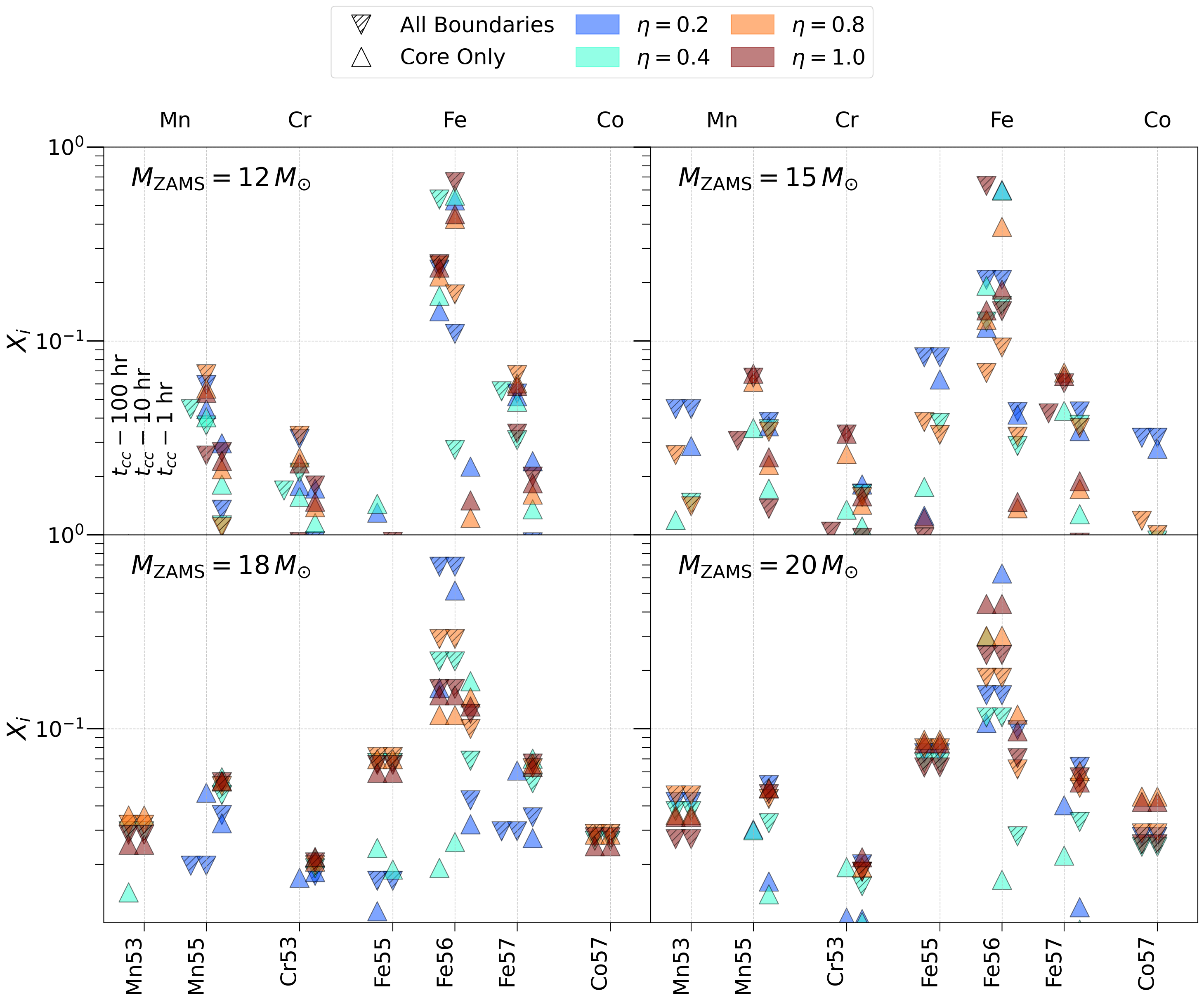}    
    \caption{The core mass fraction of ten selected isotopes for all models. For each isotope, the symbols shifted leftwards indicate the abundance at 100 hours before collapse, the central set the abundances 10 hours before collapse, and the right-shifted symbols the abundance at 1 hour before collapse.}
    \label{fig:composition}
\end{figure*}

Figure (\ref{fig:composition}) shows the composition at the center of the stars 100 hours, 10 hours, and 1 hour pre-collapse. gm(Compositions shown here are for the innermost grid zone.) Comparing models, we find that for some isotopes, the abundances differ by as much as two orders of magnitude over this time period depending upon the choice of the wind efficiency $\eta$ and overshooting treatment. In most cases the abundances show a stronger dependence on the treatment of overshooting than on mass loss, because the dispersion of the abundances is larger for stars with ``all boundaries" overshooting than with ``core only'' overshooting. Recall that in the ``all boundaries" case, we apply a modest amount of overshooting across convective boundaries outside the core. Our results suggest small changes in the treatment of overshooting may yield significant changes in a star's final composition. 
Generally, in models with ``all boundaries" overshooting, the more proton-rich isotopes have higher abundances. This is because overshooting across all convective boundaries generally raises the electron fraction of the core by mixing in material from layers farther from the core where the electron fraction is larger (see, e.g., \cite{2026MNRAS.tmp...19W}). 


\begin{figure*}
\centering
\includegraphics[width=\linewidth]{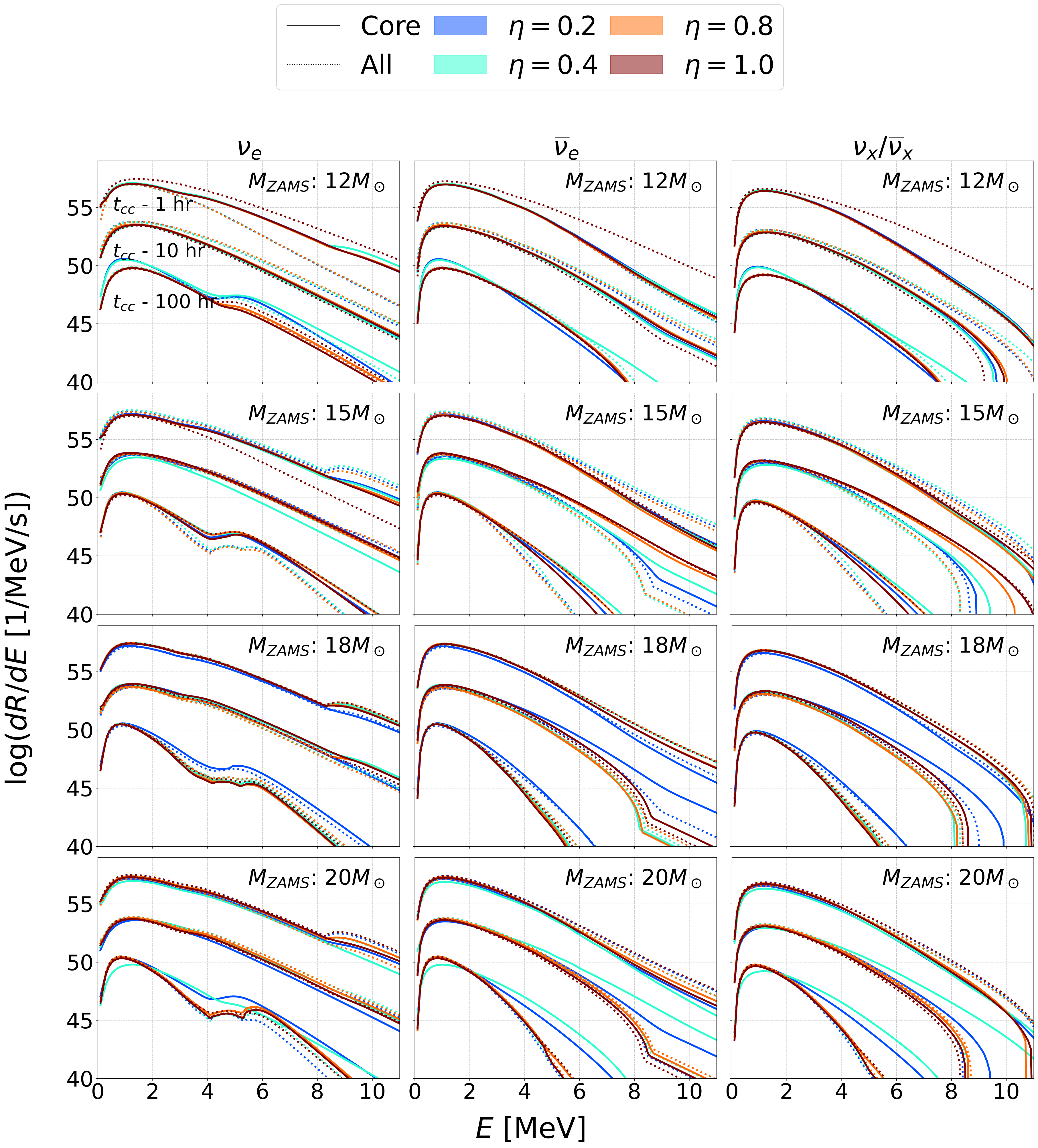}
\caption{The total neutrino emission spectra, including both beta and pair processes, for all of the RSG models. The earliest time (100 hours pre-collapse) is unshifted, while later times are shifted upward by a multiplicative factor of $10^3$ (10 hours pre-collapse) and $10^6$ (1 hour pre-collapse).  Each row contains a different mass (12, 15, 18, and 20 \Msun from top to bottom), while columns are separated into $\nu_e$, $\bar{\nu}_e$, $\nu_x$ and $\bar{\nu}_x$ from left to right.}
\label{fig:spectraAll}
\end{figure*}

\begin{figure*}
\centering
\includegraphics[width=\linewidth]{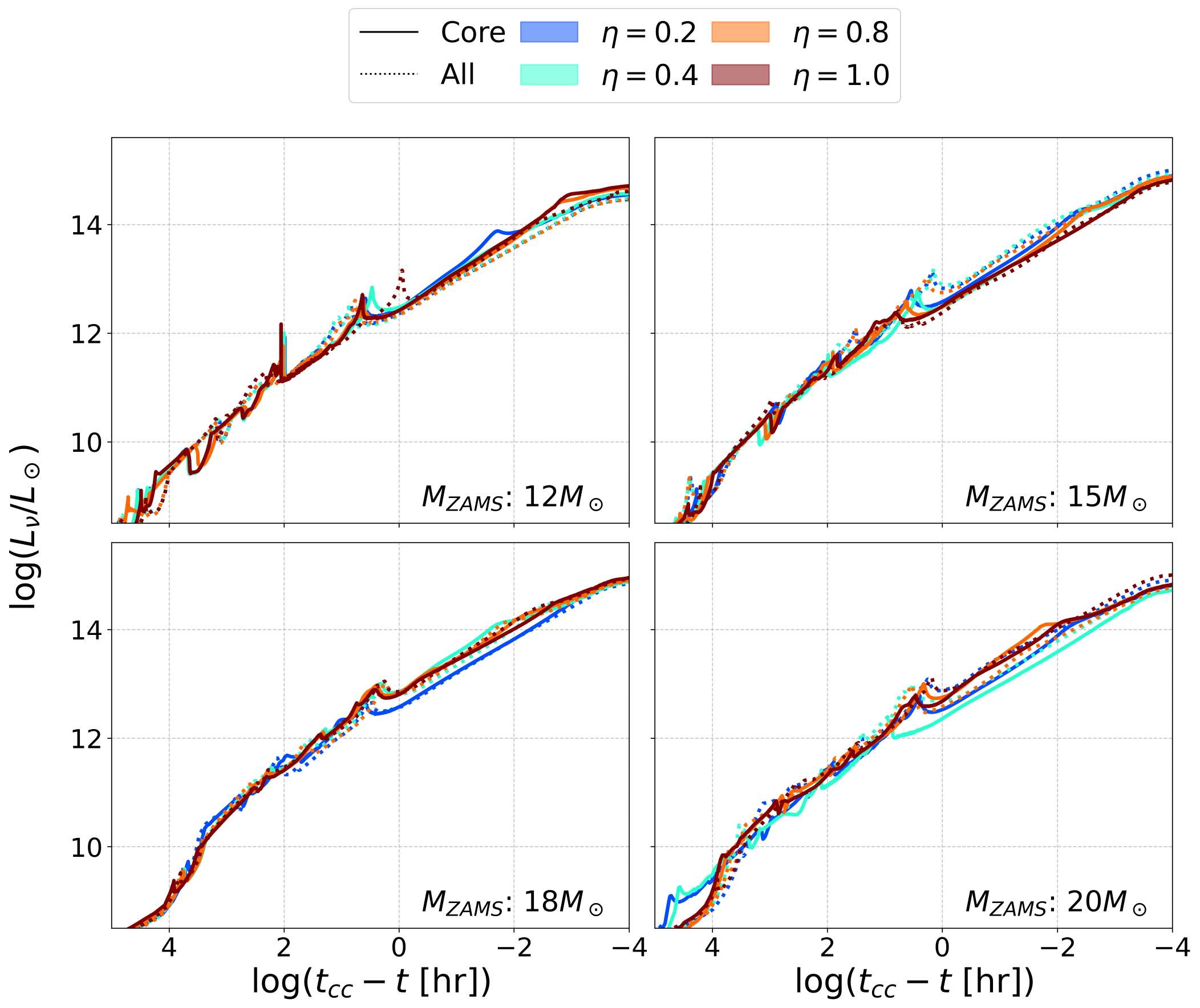}
\caption{The total neutrino luminosity, beta and pair processes combined, as a function of time for all RSG models.}
\label{fig:LnuAll}
\end{figure*}

\section{Results of the Pre-Supernova Neutrinos}
\label{sec:results2}

The variation of the core temperature, density and composition of the RSGs will be reflected in the neutrino emission from them. The time evolution of the neutrino luminosity in the 15 \Msun, $\eta = 0.8$, ``core only'' model is shown as the bottom panel of figure (\ref{fig:yeconv1508}). The neutrino luminosity is split into contributions from beta and thermal processes and shows that the neutrino luminosity generally increases over time for both processes. Changes in the structure of the star due to the initiation of nuclear burning episodes in the core or a shell lead to contemporaneous changes to the neutrino emission. Pre-supernova neutrino detection during the last $\sim$ 100 hours of the star's life with sufficient time resolution would be able to observe these changes in the core structure.  
We also see that the relative contribution of beta and thermal processes to the total changes as the star evolves. For much of the time shown, the neutrino luminosity is almost completely dominated by pair annihilation: the beta processes begin to compete around $10$ hr, and then overtake pair annihilation around $10^{-1.5}$ hr before collapse. This behavior has been observed before (see, e.g.,\ \cite{2020MNRAS.496.3961K, 2017ApJ...840....2P, 2017ApJ...848...48K}), although the domination by beta process emission has generally happened earlier in previous models. Those previous models did not include stellar mass loss, which likely accounts for the difference. We also notice that the onset of neon burning, oxygen burning, and silicon burning are all also associated with a brief burst of neutrinos from beta processes. 

The neutrino spectra as a function of energy for all 32 models are shown in figure (\ref{fig:spectraAll}).  The rows in this figure are separated by ZAMS mass, starting with \MZAMS = 12 \Msun at the top and increasing to \MZAMS = 20 \Msun at the bottom.  The columns are separated by neutrino flavor, showing $\nu_e$, $\bar{\nu}_e$, $\nu_x$ or $\bar{\nu}_x$ (the spectra of $\nu_x$ or $\bar{\nu}_x$ are equal) from left to right.  Mass loss and convective boundary treatment are indicated using the same color and line-style schemes as in earlier figures. Three times are shown in the same panels (100 hours, 10 hours, and 1 hour pre-collapse), with later times shifted upward to separate the curves.  The results for 100 hours pre-collapse are shown unshifted, while 10 hours is multiplied by $10^3$ and 1 hour is multiplied by $10^6$.  

It should be noted that we use the abundances calculated by \MESA for the beta process calculations to keep all results self-consistent.  As noted in \cite{2020MNRAS.496.3961K}, the total neutrino luminosity and even the dominant contributions to the beta spectrum can change based on the size of the network used.  While the star is roughly in nuclear statistical equilibrium (and we could compute abundances for any size network we would like with that assumption), the beta process calculations shown here are done only for the 206 isotopes that \MESA is including in the evolution for our models.

At the earliest time shown (100 hours pre-collapse), the $\nu_e$ spectrum shows many interesting features. Pair annihilation is producing more neutrinos overall at this point, but the energy for that process peaks at $E \sim 1$ MeV.  Beta processes dominate at the higher energies and add non-thermal features to the curves. In particular, all models show a shoulder in the spectrum at around $E \sim 4$ MeV, which is due to electron capture reactions on silicon and phosphorus nuclei.  The same features are not observed in the $\bar{\nu}_e$ channel because positron capture is not as prevalent.   

All of the spectra are approximately thermal by 1 hour pre-collapse. There are a few small spectral peaks seen at $E \gtrsim 8$ MeV in the $\nu_e$ channel, which are due to capture reactions on nickel isotopes. These small rises at higher energy are approximately five orders of magnitude lower than the peak, and so contribute little to the overall number of neutrinos produced. The capture peaks are smoothed out by thermal effects at even later times (see, e.g.\ \cite{ODRZYWOLEK2004303}), and have disappeared in all cases by collapse.

The dominant beta process contributions come from the isotopes expected for the burning stage of the star.  At 100 hours prior to collapse, i.e.,\ early on in the silicon burning stage, most of the neutrinos from beta processes are being produced by isotopes with $A \sim 35$, namely chlorine, sulfur, and argon.  As silicon burning progresses, a majority of the neutrinos are created from beta processes involving iron-group elements.  

The spectra in figure (\ref{fig:spectraAll}) show differences between the various RSG models.  In particular, the shape of the spectrum depends on the mass of the star, the value of $\eta$, and the convective overshooting treatment. Comparing different $\eta$ values and overshooting treatments within a single panel, we can see that the neutrino spectra can vary by several orders of magnitude at high energy for earlier times.  However, the spectra appear to converge as time progresses.  Beta processes become increasingly important as the star nears collapse, as seen in figure (\ref{fig:yeconv1508}).  These processes are highly dependent on the composition of the star, but also strongly dependent on the electron fraction $Y_e$.  Figure (\ref{fig:coreYe}) shows that the value of $Y_e$ in the core is approximately the same for all models at times less than a few hours prior to collapse, which explains the observed convergence.  

The maximum values of the spectra vary over approximately half an order of magnitude between ZAMS masses at a given time, with the \MZAMS = 12 \Msun models generally being the lowest and generally increasing as ZAMS mass increases.  As time progresses, the number of neutrinos produced increases for all stars, with the largest increases seen in the smallest stars.     
This can be seen in figure (\ref{fig:LnuAll}), which shows the total neutrino luminosity as a function of time for all RSG models.  We define the total neutrino luminosity as
\begin{equation}
    L_{\nu} = \int \left(\frac{dR_\beta}{dE_{\nu}} +  \frac{dR_{pair}}{dE_{\nu}}\right)E_{\nu} \,dE_{\nu}
\end{equation}
and divide by the photon luminosity of the Sun.  We see that over the last few hours before collapse, the neutrino luminosity increases for all stars as they approach collapse.  While it is true that the \MZAMS = 12 \Msun stars have the lowest final luminosity and the \MZAMS = 20 \Msun stars have the highest, it is not a simple monotonic relationship.  This pattern is consistent with prior studies and reflects the differences in evolution time and electron degeneracy of the various progenitors \cite{kato_pre-supernova_2015, yoshida_presupernova_2016, 2017ApJ...848...48K, 2017ApJ...840....2P}.

There are several brief spikes in the neutrino production noticeable in all models.  As seen in figure (\ref{fig:yeconv1508}), these correspond to changes in electron fraction, compactness, and the size of the convective core.  These features are present in all models, although the size, sharpness, and timing vary. Generally, the luminosity is smoother over time for the larger stars.    

All models show the same trends seen in figure (\ref{fig:yeconv1508}) as far as the contributions of pair and beta processes are concerned.  The beta process becomes dominant in the last several minutes before collapse.  This matches what has been previously observed \cite{2017ApJ...840....2P, 2017ApJ...848...48K, 2020MNRAS.496.3961K, 2020ARNPS..70..121K}, although this switch happens later in these models than in previous work.  As mentioned above, the older models did not include mass loss, and the changes in the evolution due to differences in the mass loss are likely responsible for this change.


\section{The Signal in Neutrino Detectors}
\label{sec:6}

The evolution of the neutrino emission from RSGs seen in figures (\ref{fig:spectraAll}) and (\ref{fig:LnuAll}) will lead to features in event rates and energy spectra observed in neutrino detectors here on Earth if the star is close by \cite{yoshida_presupernova_2016,2020ApJ...899..153M,PhysRevD.101.043008,Eller_2022,Machado_2022,2024ApJ...973..140A}. Fully exploring these features, their detectability, and their interpretation is a major undertaking due to large variations in detector technologies/characteristics and neutrino mixing parameters, so this will be explored in greater detail in future studies. Instead, we show here a limited sample of results to illustrate the detector features we can expect. 

We shall consider only a 20 kiloton liquid scintillator detector, i.e., the JUNO experiment, and the case where the RSG is located 200 pc from Earth. This choice of distance allows us to directly compare our calculations with those by the JUNO collaboration itself \cite{2024JCAP...01..057A}. 
To compute the event rates from our models, we use the \texttt{snewpy} software package \cite{2022ApJ...925..107B, joss}, which we have adapted by writing a custom model class to read our RSG model data (which will be released in future versions of \texttt{snewpy}), together with the \texttt{sntools} Monte Carlo event generator \cite{Migenda:2021hnl}. 
The two packages work together: \texttt{snewpy} computes the neutrino flux at Earth by extracting the neutrino spectra from the RSG model data and then applying a prescription to account for flavor transformation through the mantle of the star and decoherence; \texttt{sntools} uses these fluxes to generate events in the detector. 
Ideally those event rates would be then used in a full detector simulation to compute the observed signal. However, we do not have access to a full simulation of JUNO, so in what follows, our signals are for an idealized version of the detector that we refer to as ``JUNO-like". 
By comparing the numbers of events we calculate using this pipeline with those calculated by the JUNO collaboration \cite{2024JCAP...01..057A}, we find that detector effects (smearing, efficiency, and backgrounds) do not affect the inverse beta decay (IBD) channel as much as other channels. Thus, in what follows, we shall concentrate on the IBD event rates only. There are other detection channels in JUNO, such as the electron-neutrino elastic scattering channel, which is a channel with great potential because it has directional information that could be used to roughly locate the RSG on the sky, aiding in later efforts to identify the progenitor before shock breakout \cite{2020ApJ...899..153M}. 

\begin{figure}
    \centering
    \includegraphics[width=1.0\linewidth]{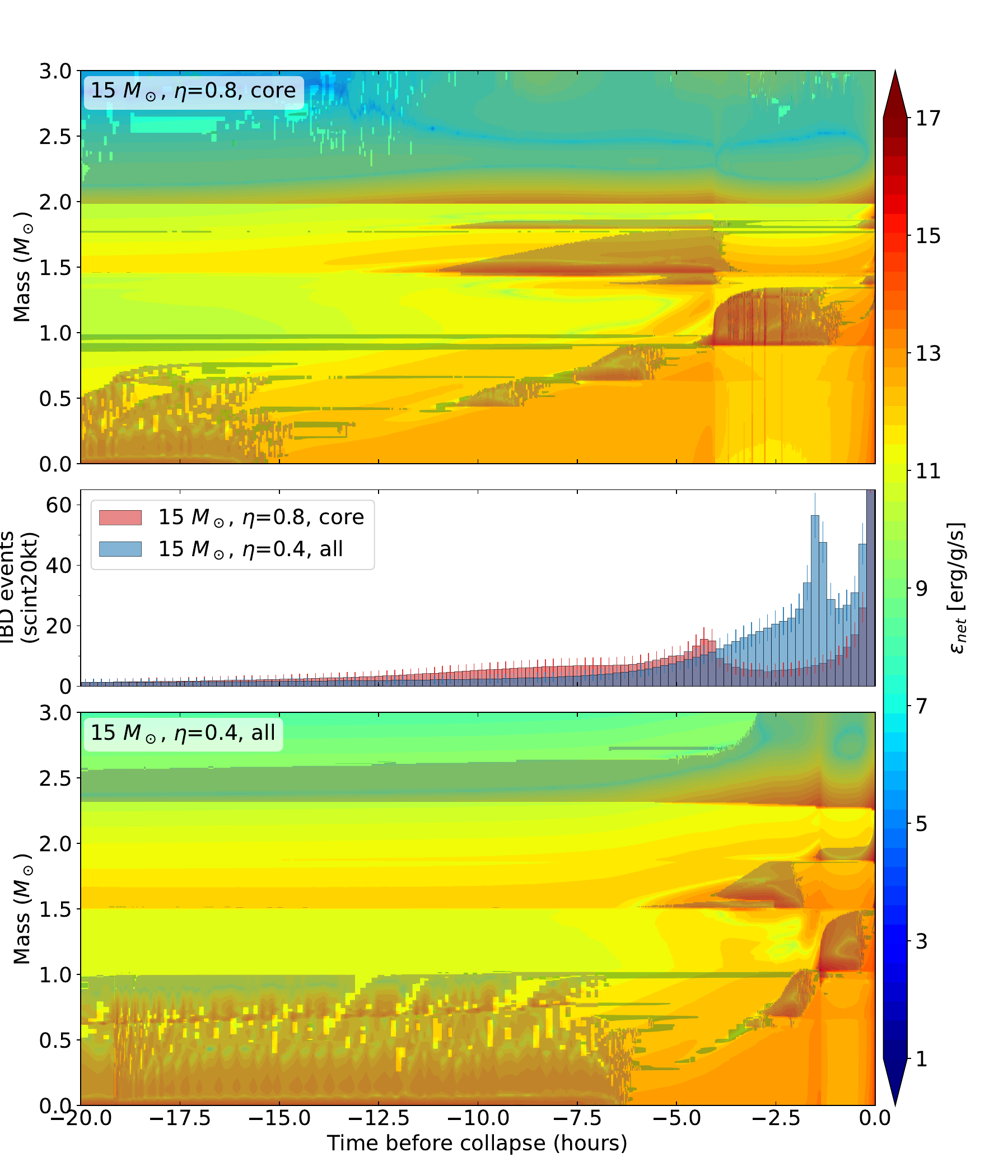}    
    \caption{The time evolution of the expected neutrino signal in a JUNO-like detector for two $\MZAMS=15\:\Msun$ models: $\eta=0.8$ ``core only'' (event counts in red), and $\eta=0.4$ all-bounds (event counts in blue). The figure shows the expected number of IBD events per 12-minute time bin. Vertical bars on each bin show the $\pm1\sigma$ Poisson uncertainty, $\sigma=\sqrt{N}$, on the expected count $N$, drawn in each model's corresponding color. Kippenhahn plots showing the nuclear energy generation rate for each model are also shown. }
    \label{fig:juno}
\end{figure}

To demonstrate that differences in the treatments of mass loss and overshooting lead to differences in the event rates, we show in the middle panel of figure (\ref{fig:juno}) a comparison of the expected number of IBD events in 12 minute bins as a function of time for two models: the $\MZAMS=15\:\Msun$, $\eta=0.8$, ``core only'' model (``Model A") and the $\MZAMS=15\:\Msun$, $\eta=0.4$, ``all boundaries" model (``Model B"). For this figure we considered only adiabatic Mikheyev-Smirnov-Wolfenstein (MSW) \cite{Mikheyev:1985aa,Mikheyev:1986tj,1978PhRvD..17.2369W} neutrino flavor transformations with a normal mass ordering and mixing parameters taken from \cite{2024JHEP...12..216E}. 
Kippenhahn plots of the nuclear energy generation rate in the two models are also shown above and below the event rates histograms so that the reader can align features in the event rates with features in the stellar models. 

Over the last hours before the collapse of the star, the general trend is that the number of IBD events in a JUNO-like detector increases exponentially. Superposed on this general trend, we observe a peak in the number of IBD events that occurs between $\sim 5$ hours and $\sim 1$ hour before collapse. 
This peak is much more prominent in Model B than Model A. A similar peak in the neutrino event rate during the pre-SN phase was seen in previous studies  \cite{yoshida_presupernova_2016,2020ApJ...899..153M}, where this peak was suggested as possibly indicating shell burning in \cite{yoshida_presupernova_2016} and core silicon ignition in \cite{2020ApJ...899..153M}. By comparing with the Kippenhahn plots, we discover this peak corresponds to changes in the shell burning of the star. 
The physical mechanism leading to the formation of a local peak in the event rates is quite simple: the neutrino emission is a function of temperature and density. The core temperature rises until shell burning ignites. After the shell ignites, the core temporarily inflates, leading to a drop in temperature and compactness---as seen in figures (\ref{fig:compactnessvstime}) and (\ref{fig:yeconv1508})---and thus a reduction in neutrino luminosity. 

\begin{figure*}
    \centering
    \includegraphics[width=1.0\linewidth]{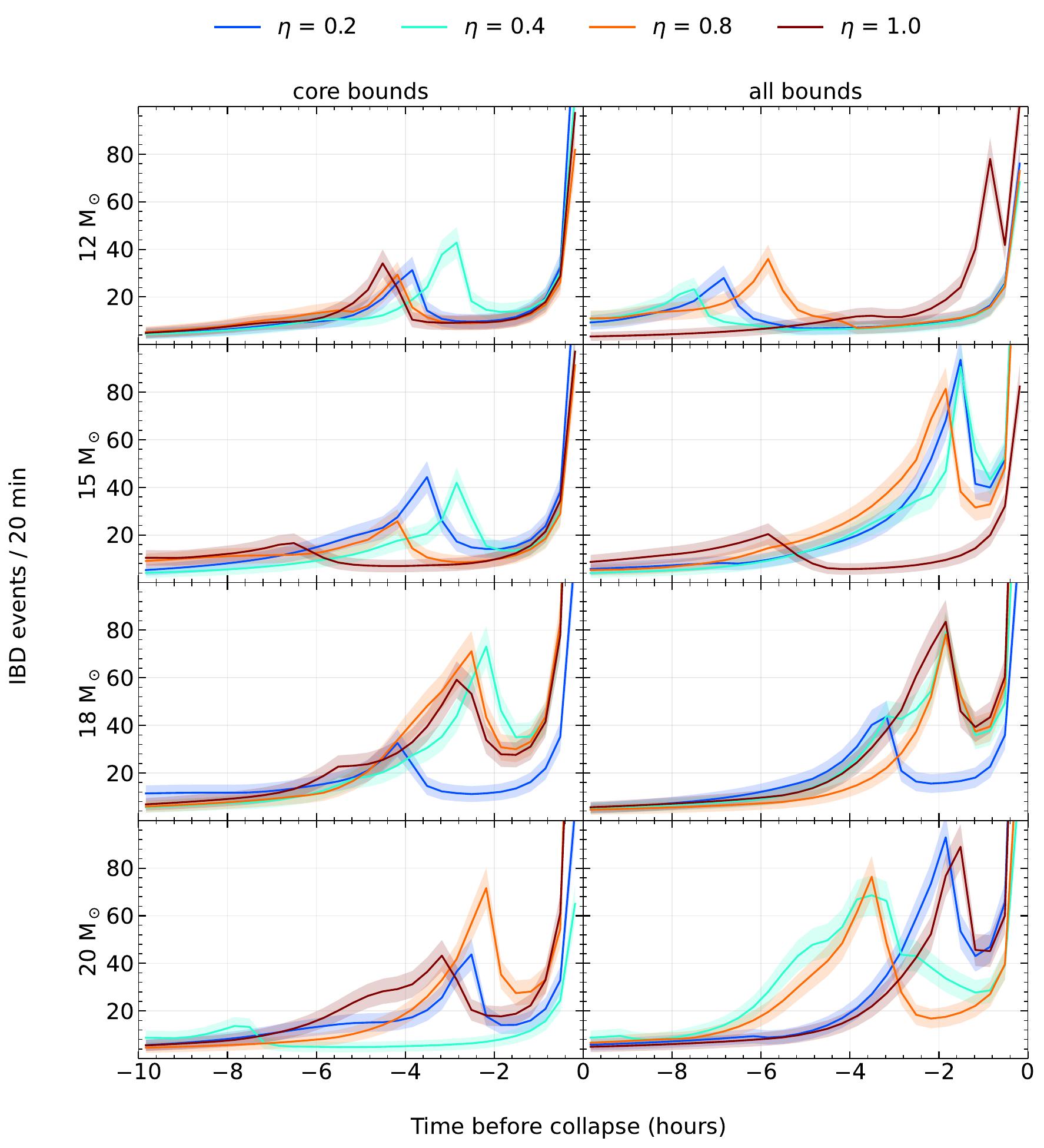}    
    \caption{The time evolution of the expected neutrino signal in a JUNO-like detector for all 32 RSG models. The figure shows the expected number of IBD events per 20-minute time bin. The shaded band around each curve is the $\pm1\sigma$ Poisson uncertainty, $\sigma=\sqrt{N}$, on the expected count $N$ in that bin, colored to match its curve.}
    \label{fig:junolike}
\end{figure*}

In figure (\ref{fig:junolike}) we show the expected IBD event rates in the JUNO-like detector for all 32 RSG models. The reader will observe that the location and height of the peak in the event rate differs between models of the same ZAMS mass and the event rates depend on the treatment of overshooting and mass loss. Peaks in the event rate occur as far as $\sim 8$ hours or as near as $\sim 1$ hour before collapse. As shown in figure (\ref{fig:juno}), these peaks correspond to the timing and strength of shell burning episodes. Comparing with figures (\ref{fig:compactnessvstime}) and (\ref{fig:compactnessatcollapse}), we discover there is a correlation between the timing of this peak and the compactness of the star at the time of collapse: in models with low compactness, the peak in IBD event rates occurs a greater amount of time before collapse. The physics behind this relationship can be understood in simple terms. The silicon shell burning will produce the last few tenths of a solar mass of ``iron'' that will cause the core of the star to collapse. The lower the compactness, the lower the density in the core, and thus the slower the silicon burns. Since the peak in the IBD event rate indicates the onset of the shell burning, for stars with lower compactness, there should be a greater amount of time between this peak and collapse, as seen in the figure.

\section{Summary \& Conclusions}\label{sec:conclusions}

In this study we have examined how different treatments of two important uncertainties in stellar evolution may affect the RSG appearance, core composition, compactness, and neutrino emission of core-collapse supernova progenitors, and the pre-supernova neutrino signal in a neutrino detector.
From the analysis of our RSG models, we find results and trends which match those observed in other studies. Like \cite{2017A&A...603A.118R}, we find the surface properties of the star do not vary greatly with the mass-loss efficiency nor the overshooting treatment for stars with $M_\mathrm{ZAMS} \lesssim 18\;\Msun$. Above this ZAMS mass, the dispersion in the effective surface temperature and luminosity begins to grow. For a fixed overshooting treatment at $M_\mathrm{ZAMS}=20\;\Msun$, the general trend is that higher mass-loss efficiency leads to lower effective temperatures and lower luminosities at the point of collapse. We similarly find that the compactness of the RSG varies considerably as it approaches the end of its life and depends sensitively on the mass-loss efficiency and overshooting treatment. The compactness is not a monotonically increasing function of time: during nuclear burning episodes in the core and shells, the compactness can decrease, i.e.,\ the core region of the star expands. This reduces the neutrino emission due to the decrease in density. Kato et al. also observed this relationship between neutrino luminosity and compactness \cite{2026arXiv260309810K}. Any convection in the core (or a shell) that occurs during an episode of burning also mixes the material within the convective region. Since the innermost regions of the star begin to deleptonize when oxygen burning begins, convection mixes material with a higher electron fraction into deleptonized regions and thus alters the isotopic composition and the beta component of the pre-SN neutrino emission. The times at which these episodes of nuclear burning and mixing occur are sensitive to the mass-loss wind efficiency $\eta$ and the treatment of overshooting. 
Finally, we demonstrated that the different treatments of overshooting and mass loss lead to differences in IBD event rates in a JUNO-like neutrino detector prior to explosion. The general trend of increasing temperature and density in the core of the star, leading to an exponentially increasing number of events, will be interrupted by shell ignitions that temporarily cause the core to expand. The shell ignitions reduce the compactness, which leads to a drop in the neutrino emission and a localized peak in the event rates $\sim 10$ hours to $\sim 1$ hour before explosion.

Our study shows that the observation of the pre-SN neutrino emission of a supernova progenitor can be used to constrain stellar evolution and supernova modeling. Inverting observed pre-SN neutrino detector event rates to infer the overshooting and mass loss will be challenging given the dependence of the RSG upon a great many other factors which we have not considered in this study. The more likely scenario is that an observed pre-supernova neutrino signal becomes a constraint upon any model of the progenitor star that produces the next nearby supernova. Modeling the explosion must begin with a progenitor with the correct structure. This constraint on the progenitor would greatly aid in testing whether our current understanding of the explosion mechanism is valid.

While this study offers valuable insights into the effect of mass loss and overshooting upon pre-SN neutrino emission, it is also not exhaustive and we propose several opportunities for further investigation.
First, we only considered a single mass-loss prescription, and varied only the scaling parameter $\eta$. There are many other mass-loss prescriptions available, and while the treatment of mass loss during the ``hot" phase is similar in many of them, in some prescriptions different physics assumptions are made for the ``cool" phase \cite{2017A&A...603A.118R}. 
Second, rotation is considered another important mixing process that affects the core composition and may even affect the final fate of the star \cite{2025MNRAS.543.2796H}. Some authors argue that rotation contributes significantly to mass loss \cite{2020MNRAS.492.5994B,2023MNRAS.524.2460B}. Our models are non-rotating. Third, metallicity has also been shown to affect neutrino luminosity, because stars with lower metallicity develop denser, hotter, and more massive cores \cite{2024ApJS..270....5F}. Our models assume solar metallicity.
Finally, we also saw in our models that the pre-SN neutrino emission was affected by nuclear shell burning physics, and the occurrence and extent of shell interactions depend strongly on mixing \cite{2019MNRAS.484.3921D}. 3D simulations suggest that dimensionality may also play an important role in shell physics, influencing nuclear energy generation and nucleosynthesis in shell mergers \cite{2025arXiv251217705I, 2020MNRAS.491..972A, 2026ApJ...997...41I}. 
Future work should explore the effects of greater variation in the treatment of overshooting, choice of mass-loss prescription, metallicity, rotation, and dimensionality, and whether such variations also lead to differences in detector event rates.

The neutrino spectra for all 32 of our RSG models are available via the \texttt{SNEWPY} model repository at \newline \url{https://github.com/SNEWS2/snewpy-models-presn}.

%% file: acknowledgments.tex
\begin{acknowledgments}

The authors are very grateful to Carla Fr\"ohlich and Frank Timmes for helpful discussions and advice. 
MAM is supported by the National Science Foundation's “Windows on the Universe: the Era of Multi-Messenger Astrophysics” Program: “WoU-MMA: Collaborative Research: A Next-Generation SuperNova Early Warning System for Multimessenger Astronomy” through Grant No. 2209449. 
CBC was supported by National Science Foundation program ``REU Site: Computational and Data Science in Astrophysics'', award No. 2150329.
JPK is supported at NC State by Department of Energy grant DE-FG02-02ER41216.
JT is supported at Oxford University by the Science and Technology Facilities Council (STFC), United Kingdom.
This work used Indiana Jetstream2 CPU at Indiana University through allocation PHY260032 from the Advanced Cyberinfrastructure Coordination Ecosystem: Services \& Support (ACCESS) program, which is supported by U.S. National Science Foundation grants \#2138259, \#2138286, \#2138307, \#2137603, and \#2138296.

\end{acknowledgments}